\documentclass[
	reprint,
	groupedaddress,
	amsmath,amssymb,
	prb,
	floatfix,
	superscriptaddress,
	column,
	aps,
]{revtex4-2}

\usepackage{amsmath}
\usepackage{bm}
\usepackage{amsfonts}
\usepackage{amssymb}
\usepackage{amsthm}
\usepackage[caption=false]{subfig}
\usepackage{graphicx}
\usepackage{hyperref}
\usepackage{cleveref}
\usepackage[T1]{fontenc}
\usepackage{libertine}
\usepackage{lineno}
\usepackage{multirow}
\usepackage{lipsum} %

\DeclareMathOperator*{\argmin}{arg\,min}

\begin{document}

\title{Constructing multicomponent cluster expansions with machine-learning and chemical embedding}
\author{Yann L. M{\"u}ller}
\affiliation{Laboratory of materials design and simulation (MADES), Institute of Materials, \'{E}cole Polytechnique F\'{e}d\'{e}rale de Lausanne}
\author{Anirudh Raju Natarajan}
\email{anirudh.natarajan@epfl.ch}
\affiliation{Laboratory of materials design and simulation (MADES), Institute of Materials, \'{E}cole Polytechnique F\'{e}d\'{e}rale de Lausanne}
\affiliation{National Centre for Computational Design and Discovery of Novel Materials (MARVEL), \'{E}cole Polytechnique F\'{e}d\'{e}rale de Lausanne}

\begin{abstract}
Cluster expansions are commonly employed as surrogate models to link the electronic structure of an alloy to its finite-temperature properties. Using cluster expansions to model materials with several alloying elements is challenging due to a rapid increase in the number of fitting parameters and training set size. We introduce the \emph{embedded cluster expansion} (eCE) formalism that enables the parameterization of accurate on-lattice surrogate models for alloys containing several chemical species. The eCE model simultaneously learns a low dimensional embedding of site basis functions along with the weights of an energy model. A prototypical senary alloy comprised of elements in groups 5 and 6 of the periodic table is used to demonstrate that eCE models can accurately reproduce ordering energetics of complex alloys without a significant increase in model complexity. Further, eCE models can leverage similarities between chemical elements to efficiently extrapolate into compositional spaces that are not explicitly included in the training dataset. The eCE formalism presented in this study unlocks the possibility of employing cluster expansion models to study multicomponent alloys containing several alloying elements.
\end{abstract}

\maketitle

\section{Introduction}
The cluster expansion (CE)\cite{sanchez1984,fontaine1994} is a versatile tool to model atomistic interactions across several material classes. On-lattice CE models are routinely used to compute  order-disorder\cite{muller2024,vanderven2018,natarajan2017a,linderalv2022}, vibrational\cite{thomas2013,thomas2014,vandewalle2002a,kadkhodaei2017} and magnetic\cite{kitchaev2021a,decolvenaere2017,decolvenaere2019,drautz2004a} thermodynamics of multicomponent materials. Though CE models primarily serve as surrogates for formation energies of atomic configurations, the method has been extended to compute tensorial properties of materials\cite{vandewalle2008}, energies of defects\cite{natarajan2020a,vanderven2001}, and to parameterize effective Hamiltonians that couple several microscopic degrees of freedom\cite{behara2024a}.

Cluster expansion models are generally trained on zero Kelvin data, such as the formation energies of a set of orderings computed with density functional theory (DFT). CE models are then used together with statistical mechanics techniques to compute finite temperature properties of materials. Within the CE formalism, the formation energy (or other material property), is expanded as a linear series of cluster basis functions multiplied with expansion coefficients. Researchers have made significant strides towards simplifying the parametrization of CE models from a limited pool of first-principles calculations. For instance, predictive CE models can be obtained by regressing against cluster functions chosen from genetic algorithms\cite{blum2005}. Conventional data science techniques such as weighting\cite{puchala2013} and cross-validation\cite{walle2002} are reported to improve the predictive power of these lattice models. More recently, regularization and cross-validation have been employed to obtain sparse cluster expansions\cite{nelson2013,mueller2009,barroso-luque2021,barroso-luque2022,zhong2022,barroso-luque2024,kadkhodaei2021}.
Efforts have also been made towards choosing training datasets that minimize the number of expensive electronic structure calculations and in characterizing errors in finite temperature properties with Bayesian techniques\cite{aldegunde2016,kristensen2014,ober2023,wen2023,chen2024d}.

Effective Hamiltonians based on the CE have provided critical insights into the behavior of structural\cite{natarajan2016,natarajan2017,smith2024,muller2024}, catalytic\cite{huang2015,cao2019}, electrochemical\cite{kitchaev2018,richards2018,lun2019,lun2021}, thermoelectric\cite{linderalv2022} and semiconductor\cite{thomas2014,zunger1994} materials. However, the use of CE models is usually limited to alloys containing $\approx$3-4 chemical species. Though there are no theoretical limitations to applying CE models to multicomponent materials, several practical difficulties arise when parameterizing and deploying multi-element CE. For instance, the number of fitting coefficients rises polynomially with the number of chemical species. As a result, parameterizing CE models for chemically-complex alloys requires large training datasets and is often computationally expensive to coarse-grain with statistical mechanics techniques. CE models have found limited applicability in predicting the finite-temperature properties of multi-principal element alloys. 
Alloys containing mixtures of several elements, also referred to as high-entropy alloys, are attractive candidates for high-performance structural materials\cite{george2019,george2020,miracle2020,miracle2024}, energy storage\cite{schweidler2024} and catalytic applications\cite{sun2021a}. Accurate atomistic models are crucial to enabling the design of the next generation of high-performance multicomponent materials.

Inspired by recent efforts in chemical dimensionality reduction, we describe a formalism to build on-lattice cluster expansion models for alloys containing several elements or site degrees of freedom. The \emph{embedded cluster expansion} (eCE), employs machine-learning techniques to simultaneously learn an embedding of site basis functions within a lower dimensional space, and the weights of an energy model. Symmetrized cluster functions constructed from the transformed site basis functions are used to compute the energy of a configuration.
We then apply the eCE formalism to build a formation energy cluster expansion for 6-component mixtures of elements in groups 5 and 6. Our results show that eCE models accurately predict formation energies in the complex alloy. Chemical trends in the alloying elements, such as the similarities between elements of the same group are naturally learnt by the model based on a small pool of electronic structure calculations. Site basis functions embedded in a three-dimensional space with the eCE model are sufficient to reproduce the energetics of the senary alloy to within 4 meV/atom. As the eCE models learn chemical similarities based on the electronic structure calculations, they are also able to extrapolate into chemical spaces that have not been sampled. As a proof of concept, we compare finite-temperature predictions of short-range order in a binary Cr-W alloy to a conventional CE and employ our new eCE framework to predict SRO in the equiatomic senary alloy.

\section{Results}
\label{sec:theory}

We begin by reviewing key aspects of the CE formalism, following which we describe a toy model to illustrate the embedded cluster expansion (eCE) before presenting the general eCE method. The eCE model is then applied to model the thermodynamics of a 6-component V-Nb-Ta-Cr-Mo-W refractory alloy.

\subsection{On-lattice cluster expansions}
Consider a crystal with $N$ sites where each site, $i$, can be occupied by $c$ chemical species (denoted $\epsilon_{1},\epsilon_{2},\cdots,\epsilon_{c}$). In general, there are $c^{N}$ distinct arrangements of the $c$ elements over the $N$ sites.
Any chemical decoration may be represented as a vector of occupation variables, $\vec{\sigma} = [\sigma_{1},\sigma_{2},\cdots,\sigma_{N}]$. The occupation variable, $\sigma_{i}$, takes a unique value for each element $\epsilon_{l}$. For instance $\sigma_{i} = l$ if $\epsilon_{l}$ occupies site $i$. The site basis functions at site $i$ can be represented as:
\begin{equation}
	\label{eq:site_function_vector}
	\vec{\varphi}(\sigma_{i}) = [\varphi_{1}(\sigma_{i}),\varphi_{2}(\sigma_{i}),\cdots,\varphi_{c}(\sigma_{i})]^{T}
\end{equation}
where $\vec{\varphi}({\sigma_{i}})$ is a vector containing $c$ site basis functions. To ensure completeness of the cluster expansion, the site functions, $\{\varphi_{1},\varphi_{2},\cdots,\varphi_{c}\}$ , must be linearly independent. Common choices for site basis functions include Chebychev polynomials\cite{sanchez1984}, occupation or indicator functions\cite{fontaine1994}, and trigonometric or sinusoidal functions\cite{vandewalle2009}. Cluster functions are then constructed by taking products of site basis functions across all $N$ sites of the crystal:
\begin{equation}
	\label{eq:cluster_functions}
	\Phi_{\vec{\alpha}}(\vec{\sigma}) = \prod_{(i,\nu)\in\vec{\alpha}} \varphi_{\nu}(\sigma_i)
\end{equation}
$\vec{\alpha}$ is a list of tuples of size $N$ with each entry containing the site index $i$ and the site function index $\nu$. Every cluster function is a product of $N$ site basis functions. It is usually convenient to choose site basis functions such that one of the functions is constant, i.e. $\phi_{1}(\sigma_{i}) = 1, \forall \sigma_{i}$. This allows for the construction of a \emph{hierarchical} cluster expansion. For example, if all site basis functions are chosen to be 1 except that of the first site, the corresponding cluster function is given by $\Phi_{1} = \phi_{2}(\sigma_{1})\times 1 \times 1 \times \cdots \times 1 = \phi_{2}(\sigma_{1})$. $\Phi_{1}$ is the cluster function associated with a point cluster located at site 1. Similarly, choosing all site functions to be 1 except for two sites will result in a cluster function that represents a \emph{pair} cluster.

Any scalar property, such as the formation energy of a configuration, is given by\cite{sanchez1984,fontaine1994}:
\begin{equation}
	\begin{split}
		E(\vec{\sigma}) =& \sum_{\vec{\alpha}\in\Lambda}\tilde{J}_{\vec{\alpha}} \Phi_{\vec{\alpha}}(\vec{\sigma})                                                                                                                                  \\
		                =& \tilde{J}_{0} + \sum_{\vec{\alpha}\in\Lambda_{point}} \tilde{J}_{\vec{\alpha}} \Phi_{\vec{\alpha}}(\vec{\sigma}) \\
                  &+ \sum_{\vec{\beta}\in\Lambda_{pair}} \tilde{J}_{\vec{\beta}} \Phi_{\vec{\beta}}(\vec{\sigma}) + \cdots
	\end{split}
	\label{eq:clex_general}
\end{equation}
where $E(\vec{\sigma})$ is the formation energy of configuration $\vec{\sigma}$, $\tilde{J}_{\vec{\alpha}}$, are \emph{effective cluster interations} (ECI) for cluster $\vec{\alpha}$, $\Lambda = \{\vec{\alpha}_{1},\vec{\alpha}_{2},\cdots\}$ is the set of all clusters in the crystal, $\Lambda_{point}$ is the set of \emph{point} clusters, and $\Lambda_{pair}$ is the set of \emph{pair} clusters. Choosing the first site basis functions to be constant partitions the total energy of a crystal into contributions arising from points, pairs, triplets etc.

\begin{figure}
	\centering
	\includegraphics[width=0.4\textwidth]{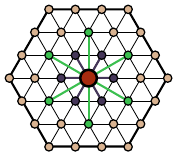}
	\caption{\textbf{Schematic illustration of a triangular lattice with two symmetrically distinct pair clusters.} The central site (shown in red) has six symmetrically equivalent nearest neighbor pair clusters marked in purple. Next-nearest neighbor pair clusters are marked in green. All pair clusters shown in a single color belong to the same orbit.}
	\label{fig:triangular_lattice_schematic}
\end{figure}

Symmetry reduces the number of independent expansion coefficients in \cref{eq:clex_general}. If cluster $\vec{\alpha}$ can be transformed to another cluster $\vec{\lambda}$ by a symmetry operation of the undecorated crystal, the interaction coefficients for both functions must be equal, i.e. $\tilde{J}_{\alpha} = \tilde{J}_{\lambda}$. All symmetrically equivalent cluster functions can be collected together in an \emph{orbit}, denoted as  $\Omega_{\vec{\alpha}} = \{\vec{\alpha},\vec{\lambda},\cdots\}$. $\vec{\alpha}$ refers to a \emph{prototype} cluster function that represents the entire orbit. The symmetrized cluster expansion contains energy contributions from each orbit:
\begin{equation}
	E(\vec{\sigma}) =\sum_{\Omega_{\vec{\alpha}} \in \Lambda} \tilde{J}_{\Omega_{\vec{\alpha}}} \sum_{\vec{\beta} \in \Omega_{\alpha}}\Phi_{\vec{\beta}}(\vec{\sigma})
	\label{eq:clex}
\end{equation}
where $\Lambda = \{\Omega_{\vec{\alpha}},\Omega_{\vec{\gamma}},\cdots\}$ is the set of all orbits. The expansion of \cref{eq:clex} can be further partitioned into energy contributions arising from each site in the crystal\cite{natarajan2018}:
\begin{equation}
	\label{eq:site_centric_clex}
	E(\vec{\sigma}) = \sum_{i=1}^{N} E_{i}(\vec{\sigma}) = \sum_{i=1}^{N} \sum_{\Omega_{\vec{\alpha}}^{i}\in \Lambda^{i}}\frac{\tilde{J}_{\Omega_{\vec{\alpha}}}}{|\vec{\alpha}|}\sum_{\vec{\delta}\in\Omega_{\vec{\alpha}}^{i}}\Phi_{\vec{\delta}}(\vec{\sigma})
\end{equation}
$E_{i}$ is the energy contributed by site $i$ to the total energy of the crystal, $\Lambda^{i}$ is the set of all clusters radiating from site $i$, $\Omega_{\vec{\alpha}}^{i}$ is the orbit of cluster function $\vec{\alpha}$ centered around site $i$ and $|\vec{\alpha}|$ is the number of sites in the cluster. The additional factor $\frac{\tilde{J}_{\Omega_{\vec{\alpha}}}}{|\vec{\alpha}|}$ arises due to overcounting of each cluster in the site-centric cluster expansion. For simplicity of notation we will define $J_{\Omega_{\vec{\alpha}}} = \frac{\tilde{J}_{\Omega_{\vec{\alpha}}}}{|\vec{\alpha}|}$. The symmetrized cluster functions $\Theta_{\Omega_{\vec{\alpha}}^{i}} = \sum_{\vec{\delta}\in\Omega_{\vec{\alpha}}^{i}}\Phi_{\vec{\delta}}(\vec{\sigma})$ are site-centric descriptors that can distinguish between all symmetrically distinct neighborhoods around site $i$. \Cref{fig:triangular_lattice_schematic} schematically shows the orbit of nearest neighbor and next-nearest neighbor pair clusters on a triangular lattice. Each cluster will additionally contain an orbit of cluster functions. The symmetrized site-centric descriptors can serve as inputs to any regression model that parameterizes the site energy $E_{i}$\cite{natarajan2018}:
\begin{equation}
	\label{eq:site_energy_expansion}
	E(\vec{\sigma}) = \sum_{i=1}^{N} E_{i}(\{\Theta_{\Omega_{\vec{\alpha}}},\Theta_{\Omega_{\vec{\beta}}},\cdots\})
\end{equation}
Although there are an infinite number of site-centric descriptors, in practice, the number of cluster functions must be truncated. Linear CE models typically enumerate cluster functions up to a maximal cluster size and cluster radius before fitting the CE model with techniques such as compressed sensing, genetic algorithms etc. Non-linear CE models may be advantageous as they have been found to converge at significantly smaller cluster sizes than linear models\cite{natarajan2018}.

\begin{figure}[h]
	\centering
	\includegraphics[width=0.4\textwidth]{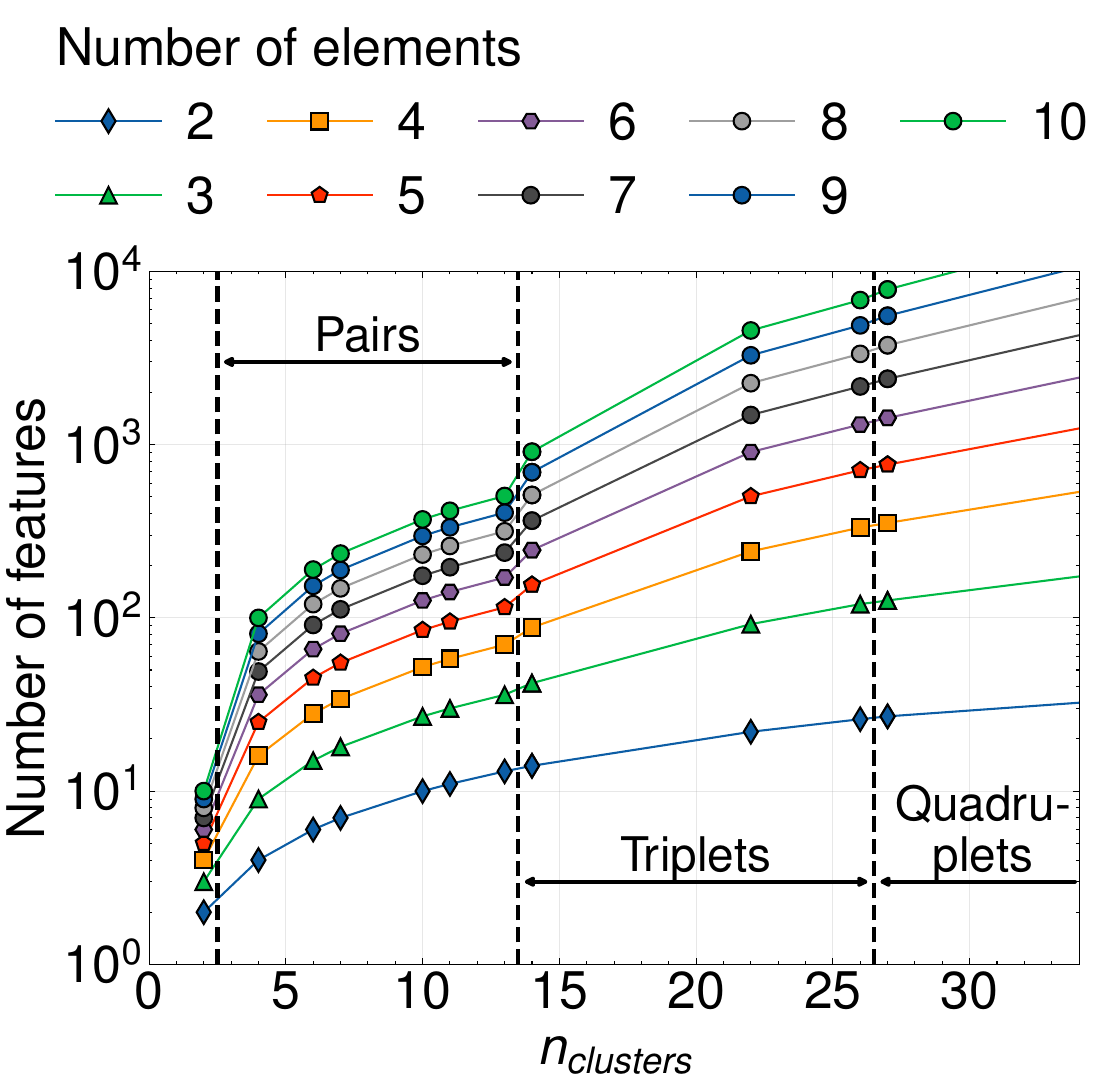}
	\caption{\textbf{Polynomial increase in the number of features with the number of alloying elements.} Variation in the number of symmetrically distinct cluster functions with the number of unique clusters in multicomponent alloys containing between 2 and 10 alloying elements. Clusters are enumerated on a bcc crystal structure with a maximum size of 10\AA$\,$ for the pair clusters and 7\AA$\,$ for the triplet clusters.}
	\label{fig:nbre_features}
\end{figure}

Though exact and complete, the cluster expansion formalism is not easily amenable to capture atomistic interactions in multicomponent alloys. The number of cluster functions in a cluster containing $k$ sites with $c$ chemical species scales polynomially as $(c-1)^{k}$. As shown in \cref{fig:nbre_features}, in alloys with over 5 chemical species, the number of features can exceed a thousand even for relatively small cluster sizes. The rapid growth in the number of cluster functions with chemical complexity is thus a major impediment to parameterizing accurate multicomponent cluster expansions.

\subsection{Encoding chemical similarity through a linear transformation of site basis functions}
\label{sec:encod-chem-simil}

Multicomponent alloys often have additional degeneracies that arise due to chemical similarities between elements. This is typically manifested as relationships between interaction coefficients of the multicomponent CE. For instance, consider a ternary alloy with three chemical elements A, B and C that can occupy each site of the triangular lattice (\cref{fig:triangular_lattice_schematic}). If the elements B and C are chemically similar, we would expect that the ECI of cluster functions that involve either the B or C element are related. This is readily seen in a CE that employs the occupation basis. Occupation site basis functions adopt the following values:
\begin{equation}
	\boldsymbol{\varphi} = \begin{bmatrix}
		\varphi_1 (A) & \varphi_1 (B) & \varphi_1 (C) \\
		\varphi_2 (A) & \varphi_2 (B) & \varphi_2 (C) \\
		\varphi_3 (A) & \varphi_3 (B) & \varphi_3 (C) \\
	\end{bmatrix} =
	\begin{bmatrix}
		1 & 1 & 1 \\
		0 & 1 & 0 \\
		0 & 0 & 1
	\end{bmatrix}
	\label{eq:site_basis_functions_ternary}
\end{equation}
The matrix $\boldsymbol{\varphi}$ contains the value of the three site basis functions $\varphi_{1},\varphi_{2},\varphi_{3}$. The columns of the matrix correspond to the values of the basis functions if either A, B or C occupies the site. Assuming that nearest neighbor pair interactions are sufficient to describe the ordering energies in the ternary alloy, the formation energy of a configuration is given by:
\begin{equation}
	\label{eq:ternary_clex}
	\begin{split}
		E(\vec{\sigma}) & = NJ_{0} + J_{B} \sum_{i}\varphi_{2}(\sigma_{i}) + J_{C} \sum_{i}\varphi_{3}(\sigma_{i})                                          \\
		                & + J_{BB}^{NN} \sum_{i,j \in NN} \varphi_{2}(\sigma_{i}) \varphi_{2}(\sigma_{j})                                                   \\
		                & + J_{CC}^{NN} \sum_{i,j \in NN} \varphi_{3}(\sigma_{i}) \varphi_{3}(\sigma_{j})                                                   \\
		                & + J_{BC}^{NN} \sum_{i,j \in NN} (\varphi_{2}(\sigma_{i}) \varphi_{3}(\sigma_{j})+\varphi_{2}(\sigma_{j}) \varphi_{3}(\sigma_{i}))
	\end{split}
\end{equation}
where $J_{0}$ is the energy of the empty cluster, $J_{B},J_{C}$ are the point energies of the B and C chemical elements, $J_{BB}^{NN},J_{CC}^{NN},J_{BC}^{NN}$ are the pair energies of a B-B, C-C and B-C pair respectively.
The chemical similarity of B and C should manifest in the energy expansion as equalities between the interaction coefficients. Specifically, $J_{B} = J_{C}$ and $J_{BB}^{NN} = J_{CC}^{NN} = J_{BC}^{NN}$. In practice, the relationships between ECI are learnt based on a training dataset of electronic structure calculations, but are never exploited. 

The degeneracies between ECI for this system can be used to learn a simpler CE. Consider the following linear transformation of the site basis functions of \cref{eq:site_basis_functions_ternary}:
\begin{equation}
	\label{eq:site_basis_functions_ternary_projected}
	\begin{split}
		\mathcal{T} \vec{\varphi}(\sigma_{i}) & = \vec{\tilde{\varphi}}(\sigma_{i}) \\
		\begin{bmatrix}
			1 & 0 & 0 \\
			0 & 1 & 1
		\end{bmatrix}
		\begin{bmatrix}
			\varphi_{1}(\sigma_{i}) = 1 \\
			\varphi_{2}(\sigma_{i})     \\
			\varphi_{3}(\sigma_{i})
		\end{bmatrix}        & =
		\begin{bmatrix}
			1 \\
			\varphi_{2}(\sigma_{i}) + \varphi_{3}(\sigma_{i})
		\end{bmatrix}
	\end{split}
\end{equation}
\begin{figure}[h]
	\centering
	\includegraphics[width=0.3\textwidth]{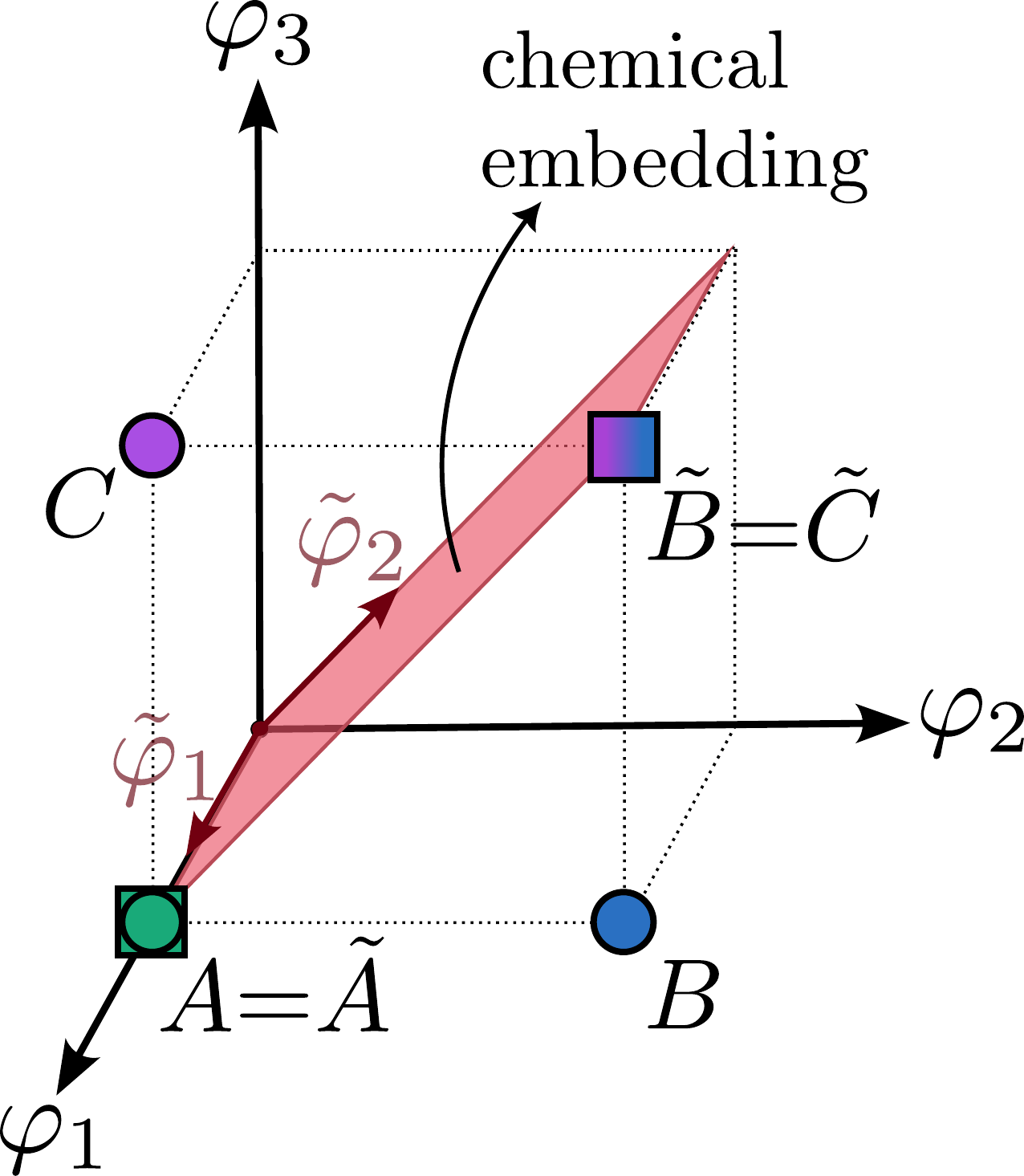}
    \caption{\textbf{Effect of the embedding matrix on site basis functions.} The depicted embedding matrix corresponds to the transformation in \cref{eq:site_basis_functions_ternary_projected}. The values of the site basis functions evaluated for A, B and C are shown as solid circles. The original 3-dimensional space is transformed into a two-dimensional sub-space (shown in red). The values of the site basis functions, evaluated in the transformed basis for the B and C chemical specie are the same. Site basis function values for the A specie remain unchanged.}
	\label{fig:projection_schematic}
\end{figure}
where the original site basis functions, $\vec{\varphi}(\sigma_{i})$, have been transformed by the matrix $\mathcal{T}$, into a set of new site basis functions $\vec{\tilde{\varphi}}(\sigma_{i})$. The transformed site basis functions, $\vec{\tilde{\varphi}}(\sigma_{i})$ span a two-dimensional sub-space within the three-dimensional space of the original site functions. The projected space is embedded as shown in \cref{fig:projection_schematic}. The new site functions are not mathematically complete and evaluate to identical values for both the B and C specie:
\begin{equation}
	\boldsymbol{\tilde{\varphi}} = \begin{bmatrix}
		\tilde{\varphi}_1 (A) & \tilde{\varphi}_1 (B) & \tilde{\varphi}_1 (C) \\
		\tilde{\varphi}_2 (A) & \tilde{\varphi}_2 (B) & \tilde{\varphi}_2 (C) \\
	\end{bmatrix} =
	\begin{bmatrix}
		1 & 1 & 1 \\
		0 & 1 & 1
	\end{bmatrix}
	\label{eq:site_basis_functions_ternary_projected_values}
\end{equation}
In the site basis function space of $\vec{\tilde{\varphi}}$, it is impossible to distinguish between the B and C specie. However, a CE constructed through tensor products of the transformed site basis functions, $\vec{\tilde{\phi}}$ can reproduce the ordering energies of the three chemical species on a triangular lattice:
\begin{equation}
	\label{eq:ternary_clex_projected}
	\begin{split}
		E(\vec{\sigma}) = & NJ_{0} + J_{B} \sum_{i=1}^{N}\tilde{\varphi}_{2}(\sigma_{i})                                       \\
		                  & + J_{BB}^{NN} \sum_{i,j \in NN}^{N}\tilde{\varphi}_{2}(\sigma_{i}) \tilde{\varphi}_{2}(\sigma_{j})
	\end{split}
\end{equation}
As the transformed site basis functions evaluate to identical values for both the B and C specie, the CE of \cref{eq:ternary_clex_projected} will compute the exact same energy for orderings where B atoms are replaced with C or vice-versa. Embedding the ``chemical similarity'' of elements B and C into the site basis functions allowed us to reduce the number of cluster basis functions from 6 to 3. Though chemical rules are often not known \emph{a-priori}, they can be simultaneously learnt together with the interaction coefficients based on a small pool of electronic structure calculations.

\subsection{Embedded Cluster Expansions (eCE)}
\label{sec:mult-clust-expans}

In general, any set of linearly independent and complete site basis functions can be embedded in a lower dimensional space through a linear transformation:
\begin{equation}
	\label{eq:site_basis_function_projection}
	\vec{\tilde{\varphi}} (\sigma_{i}) = \mathcal{T} \vec{\varphi} (\sigma_{i})
\end{equation}
$\vec{\varphi} (\sigma_{i})$ is a vector of size $c\times 1$, the transformed site basis $\vec{\tilde{\varphi}}(\sigma_{i})$ is a vector of size $k\times 1$ and the transformation $\mathcal{T}$ is a matrix of size $k\times c$, where $k\le c$ and the rank of the transformation matrix is $k$. The elements of $\mathcal{T}$ linearly mix the site functions of $\vec{\varphi}$ to obtain a lower-dimensional site basis. Maintaining the hierarchy of the cluster expansion requires that the constant site basis function remains in the transformed site basis. In practice, this can be enforced by fixing the first row of $\mathcal{T}$ to $[1,0,0 \cdots]$. The remaining entries of the transformation are learnable parameters of the model.

Cluster functions at each site are computed from the transformed site basis functions similar to \cref{eq:cluster_functions} and symmetrized as detailed in \cref{eq:site_centric_clex}:
\begin{equation}
	\label{eq:symmetrized_projected_cluster_function}
	\tilde{\Theta}_{\Omega^{i}_{\alpha}} = \sum_{\vec{\delta}\in\Omega_{\alpha}^{i}}\prod_{(j,\nu)\in\vec{\delta}}\tilde{\varphi}_{\nu}(\sigma_{j})
\end{equation}

Site-centric energies can then be expanded in terms of the symmetrized cluster functions:
\begin{equation}
	\label{eq:site_centric_projected_energies}
	E(\vec{\sigma}) = \sum_{i=1}^{N}E_{i}(\{\tilde{\Theta}_{\ \Omega^{i}_{\alpha}},\cdots\})
\end{equation}
The symmetrized cluster functions are invariant under all symmetry operations of the disordered phase\cite{natarajan2018} and can be used as inputs to any regression method to parameterize $E_{i}$. 

We refer to the CE formalism that embeds the site basis functions in a lower dimensional space as \emph{embedded cluster expansions} or \emph{eCE}. Throughout this study, we will indicate the number of effective chemical elements, i.e. the number of rows in $\mathcal{T}$ with a number. For instance, 2-eCE refers to a model with the site functions embedded in a two-dimensional space. Though the model is mathematically incomplete, exploiting chemical similarities between the elements significantly reduces the complexity of the energy expansion and results in fewer site-centric descriptors. For instance, 2-eCE models have an identical number of descriptors as a binary cluster expansion. A 2-eCE model of a 6-component alloy with all clusters shown in \cref{fig:nbre_features} will require $\approx 10^{1}$ features, while the exact CE will contain $\approx 10^{3}$ descriptors. 

The elements of the transformation matrix and regression coefficients for the energy model can be simultaneously learnt by minimizing a loss function through gradient descent techniques:
\begin{equation}
	\label{eq:loss_function_training}
	\mathcal{L} = \argmin_{\boldsymbol{w},\mathcal{T}} {\sum_{\vec{\sigma}} \Big( E^{DFT}(\vec{\sigma}) - \sum_{i=1}^{N} E^{eCE}_{i}(\vec{\sigma},\boldsymbol{w},\mathcal{T}) \Big)^{2} + \mathcal{L}_{reg}}
\end{equation}
where the first term is a least-squares error between the computed formation energies and the values predicted by the model, $\mathcal{L}_{reg}$ is a regularization term to prevent overfitting, $\boldsymbol{w}$ are the energy weights and $\mathcal{T}$ is the transformation matrix.

We demonstrate the advantages of the eCE formalism in a senary V-Nb-Ta-Cr-Mo-W alloy. This senary alloy is of current interest for high-temperature applications\cite{george2020,george2019,miracle2020,miracle2024}. Mixtures of elements in groups 5 and 6 of the periodic table form disordered solid solutions on the body-centered cubic crystal structure, or ordered Laves phases\cite{natarajan2020}.  We focus on the configurational thermodynamics of orderings on the bcc crystal structure for the rest of this study.

\subsection{Hyperparameter optimization}

\label{sec:hyperp-optim}
\begin{figure}[h]
	\centering
	\includegraphics[width=0.45\textwidth]{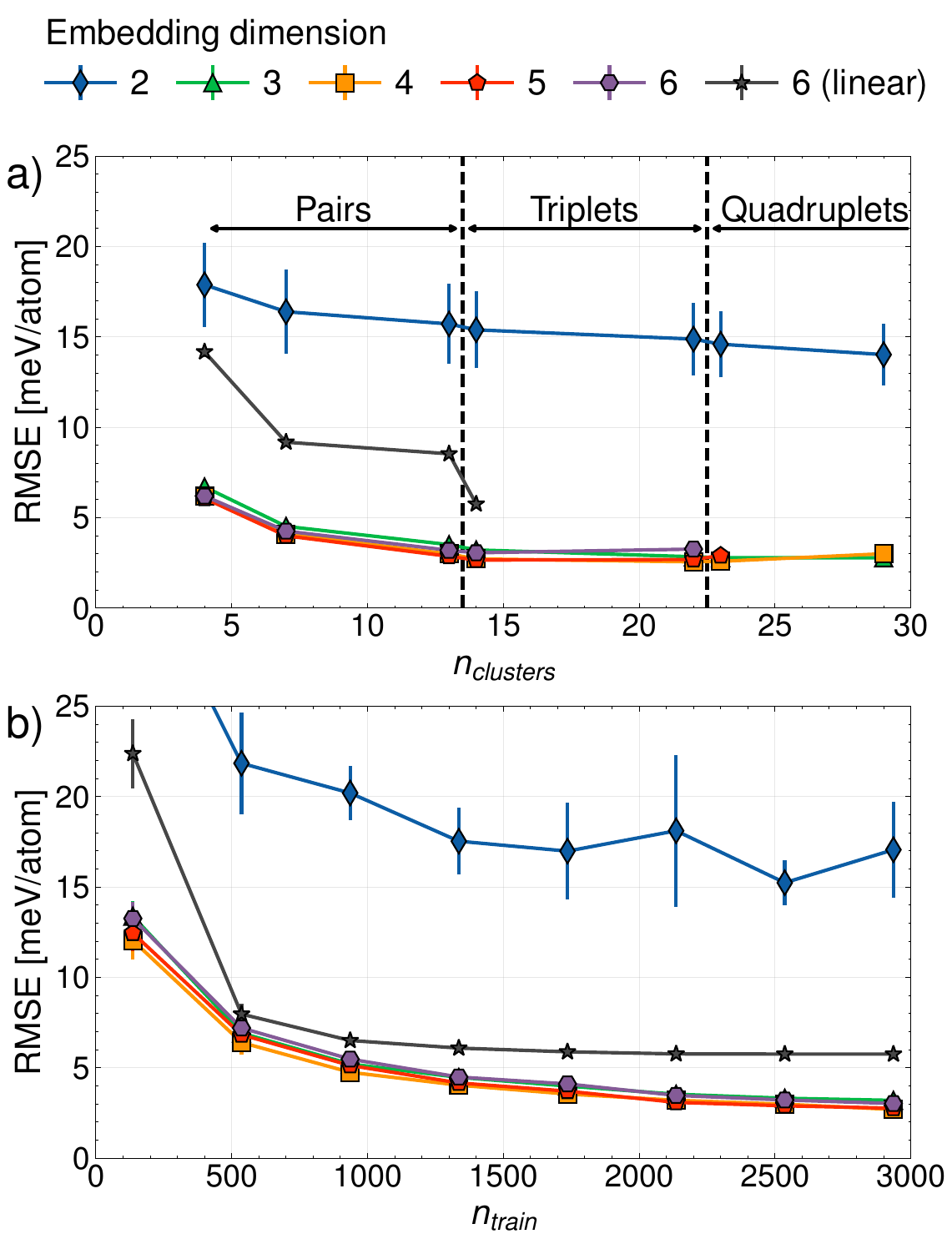}
	\caption{\textbf{Learning curves for eCE and conventional CE.} Average validation errors (solid markers) and the standard deviations (error bars) are computed over 10 separate model parameterizations. (a) Variation in the test error with the number of clusters for training and test datasets containing 2936 and 1147 configurations, respectively. (b) Variation in the test error with the number of training datapoints for eCE models with 14 clusters.}
	\label{fig:hyperparameter_optimization}
\end{figure}

Parameterizing embedded cluster expansions (eCE) with \cref{eq:site_basis_function_projection,eq:symmetrized_projected_cluster_function,eq:site_centric_projected_energies} requires several hyperparameters to be carefully tuned. \Cref{fig:hyperparameter_optimization} shows the variation in validation error with the number of effective chemical species in the eCE (i.e. the embedding dimension), the number of symmetrically distinct clusters, and the number of datapoints in the training dataset. The CE models of \cref{fig:hyperparameter_optimization}a are trained to reproduce formation energies of 2936 randomly sampled datapoints. Separate models were parameterized over 10 random instantiations of the training dataset. The site energies of eCE models are computed with neural networks. Additional information about the neural network parameters, and regularization are provided in \cref{sec:methods}. 

The mean validation error in \cref{fig:hyperparameter_optimization}a is computed over 1147 configurations that are not included in the training set. A cluster expansion containing two effective chemical species, 2-eCE, results in high validation errors of $\approx 15$meV/atom. As 2-eCE embeds the 6 linearly independent site basis functions in a two-dimensional sub-space, this model is similar to a ``binary'' cluster expansion. Unlike a conventional binary cluster expansion where one of the site functions can take two possible values, the site function in 2-eCE can take six distinct values. The saturation in validation error with increasing number of clusters suggests the model with two site basis functions lacks the chemical flexibility to reproduce the ordering energies of this senary alloy. Increasing the dimensionality of the projected site basis functions to three lowers the validation error to $\approx 3$ meV/atom. In fact, the ``ternary'' 3-eCE model is essentially as accurate as cluster expansions with six linearly independent site basis functions (red line in \cref{fig:hyperparameter_optimization}a). The relatively small spread in validation errors for eCE models suggests that the models are not very sensitive to the exact configurations included in the training dataset.

\Cref{fig:hyperparameter_optimization}a also compares a linear CE model (parameterized with ridge regression) to non-linear models that use neural networks. Similar to a previous study\cite{natarajan2018}, the linear model requires more cluster basis functions to reach accuracies comparable to non-linear models. Parameterizing a linear cluster expansion model with triplet cluster sizes larger than 4$\text{\AA}$ was prohibitively expensive due to the large memory requirements needed for ridge regression. Cluster functions built with pair clusters up to a size of 10$\text{\AA}$ and triplet clusters with a size of 4$\text{\AA}$ are found to be highly accurate for non-linear models. Linear eCE models are found to have significantly higher prediction errors than non-linear eCE models that employ neural networks to compute site energies.

Having identified the optimal number of clusters, we next study the variation in validation error with the size of the training dataset. \Cref{fig:hyperparameter_optimization}b shows learning curves of eCE models with projection dimensions ranging from 2-6. 2-eCE models trained with up to 3000 datapoints have validation errors of over 15 meV/atom. Interestingly, $\approx 500-1000$ randomly chosen training datapoints are sufficient to reproduce the complex chemical interactions in the senary alloy for eCE models that contain 3 or more embedding dimensions. The validation errors for a linear senary CE are higher than the non-linear models.

\begin{figure}[h]
	\centering
	\includegraphics[width=0.41\textwidth]{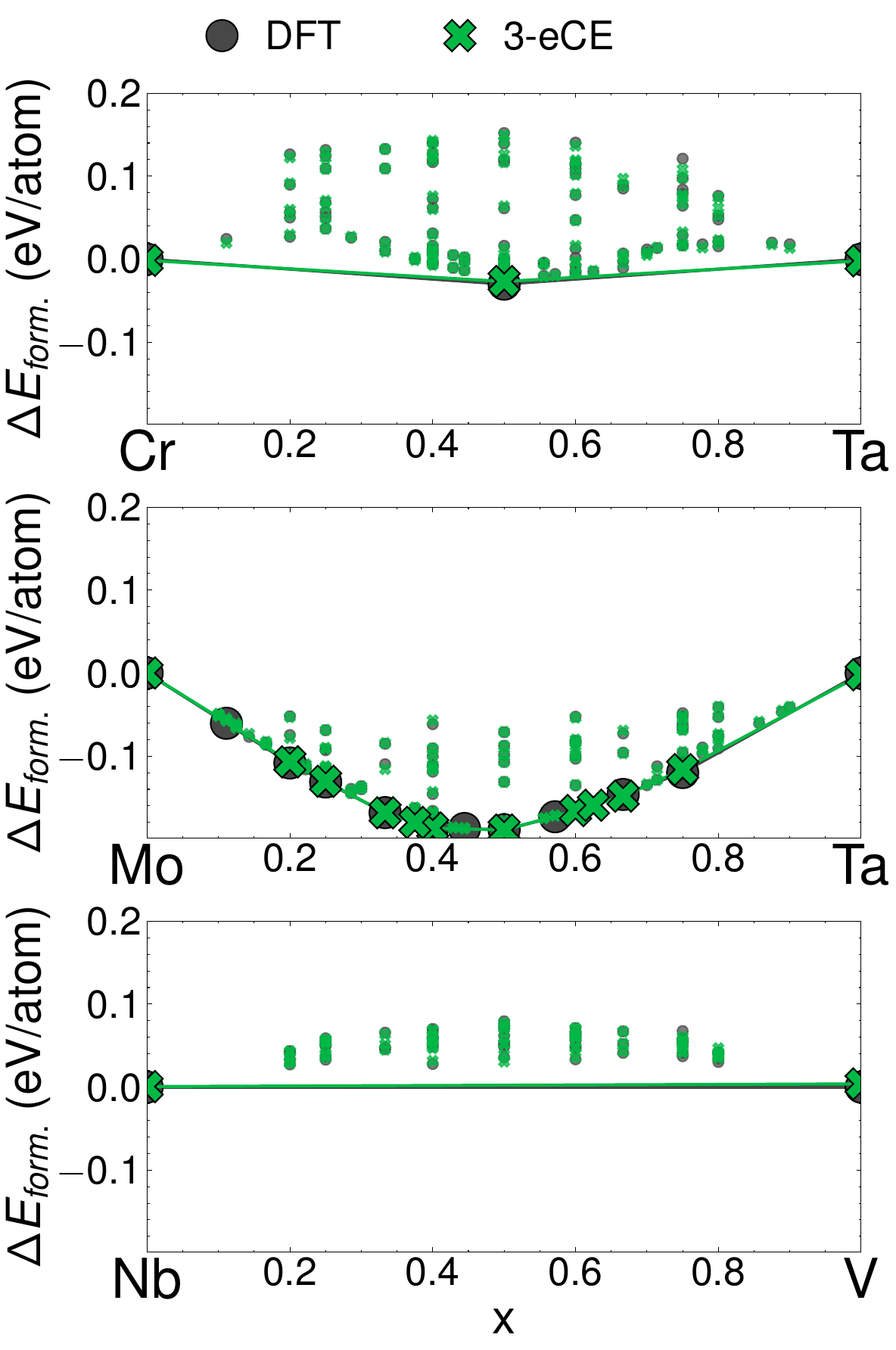}
	\caption{\textbf{Formation energies in the Cr-Ta, Mo-Ta and Nb-V binary alloys.} Grey circles are formation energies computed with DFT and green crosses with 3-eCE. Configurations on the convex hull are shown with larger markers. Convex hulls for each system are shown with either a green line (3-eCE) or a grey line (DFT).}
	\label{fig:convex_hull_compare}
\end{figure}

Often, despite low validation errors, atomistic models may fail to reproduce phase stability at low temperatures. \Cref{fig:convex_hull_compare} depicts the formation energies of orderings in three binary alloys. DFT calculations (grey circles in \cref{fig:convex_hull_compare}) predict a single binary ground state in Cr-Ta, no stable ground states in the Nb-V alloy, and several stable ground states in the Mo-Ta alloy system. A 3-eCE model trained with 2936 datapoints and 14 symmetrically distinct clusters reproduces all the salient features of the zero Kelvin energies in binary alloys (green crosses in \cref{fig:convex_hull_compare}). The stable states in Cr-Ta and Nb-V are exactly reproduced by the 3-eCE model, while most ground states are captured by the 3-eCE in Mo-Ta. Minor discrepancies between eCE and the DFT calculations may be due to fitting errors or small numerical errors in the electronic structure calculations. Nevertheless, there is excellent agreement between the 3-eCE model and DFT, indicating that eCE models are able to capture thermodynamic ground states in addition to the overall ordering energetics of multicomponent alloys.

\subsection{Extrapolating in chemical space}
\label{sec:extr-chem-space}

\begin{figure}[h]
	\centering
	\includegraphics[width=0.41\textwidth]{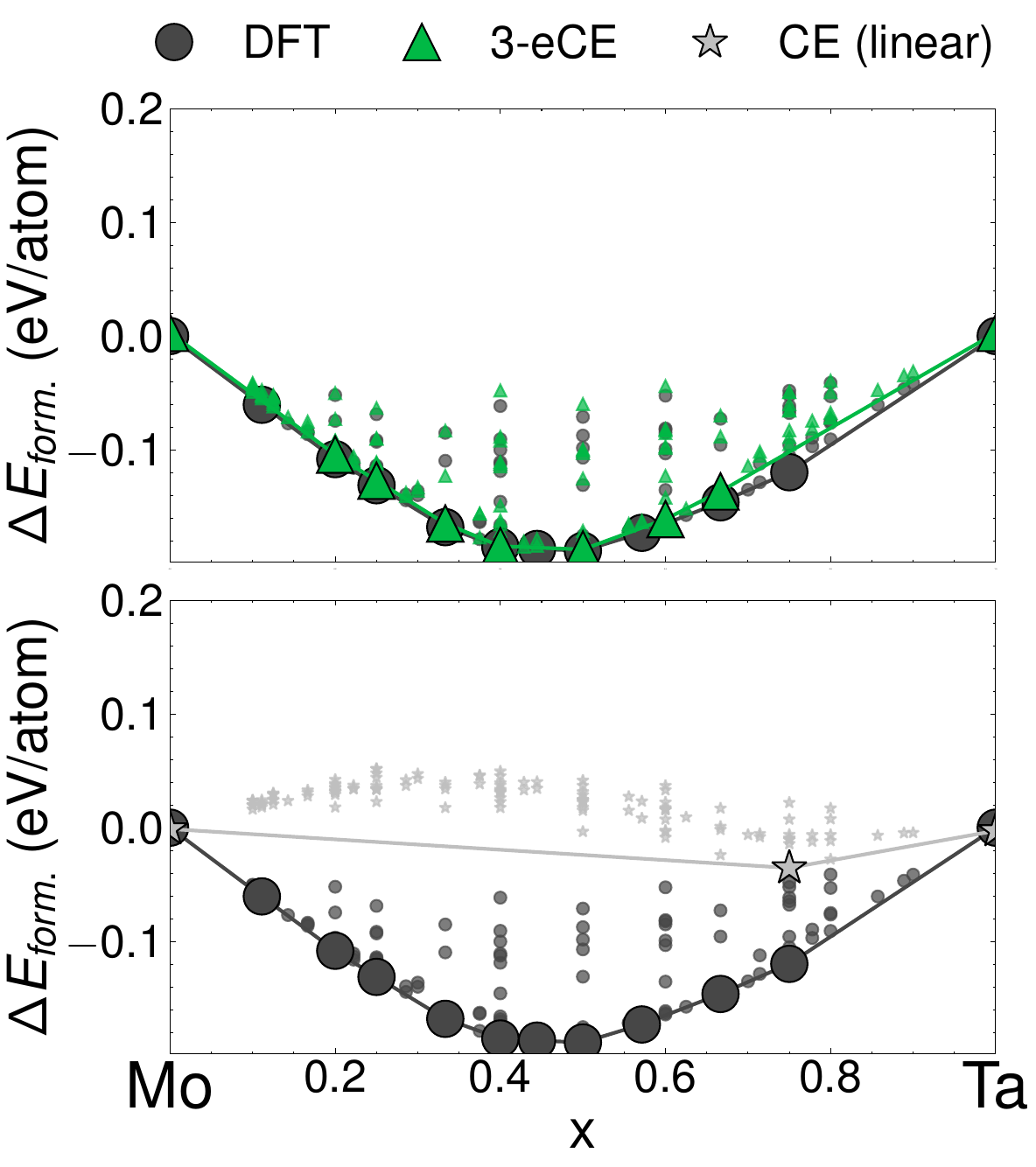}
	\caption{\textbf{Chemical extrapolation.} Formation energies of orderings in the Mo-Ta binary alloy as computed with DFT (grey circles), a 3-eCE model (green triangles) and a linear CE (grey stars). The 3-eCE and CE models are trained on a dataset that does not contain any configurations with both Mo and Ta. Convex hulls are shown as dark grey (DFT), green (3-eCE) and light grey (linear CE) lines.}
	\label{fig:binary_dropout_example}
\end{figure}

\begin{figure}[h]
	\centering
	\includegraphics[width=0.47\textwidth]{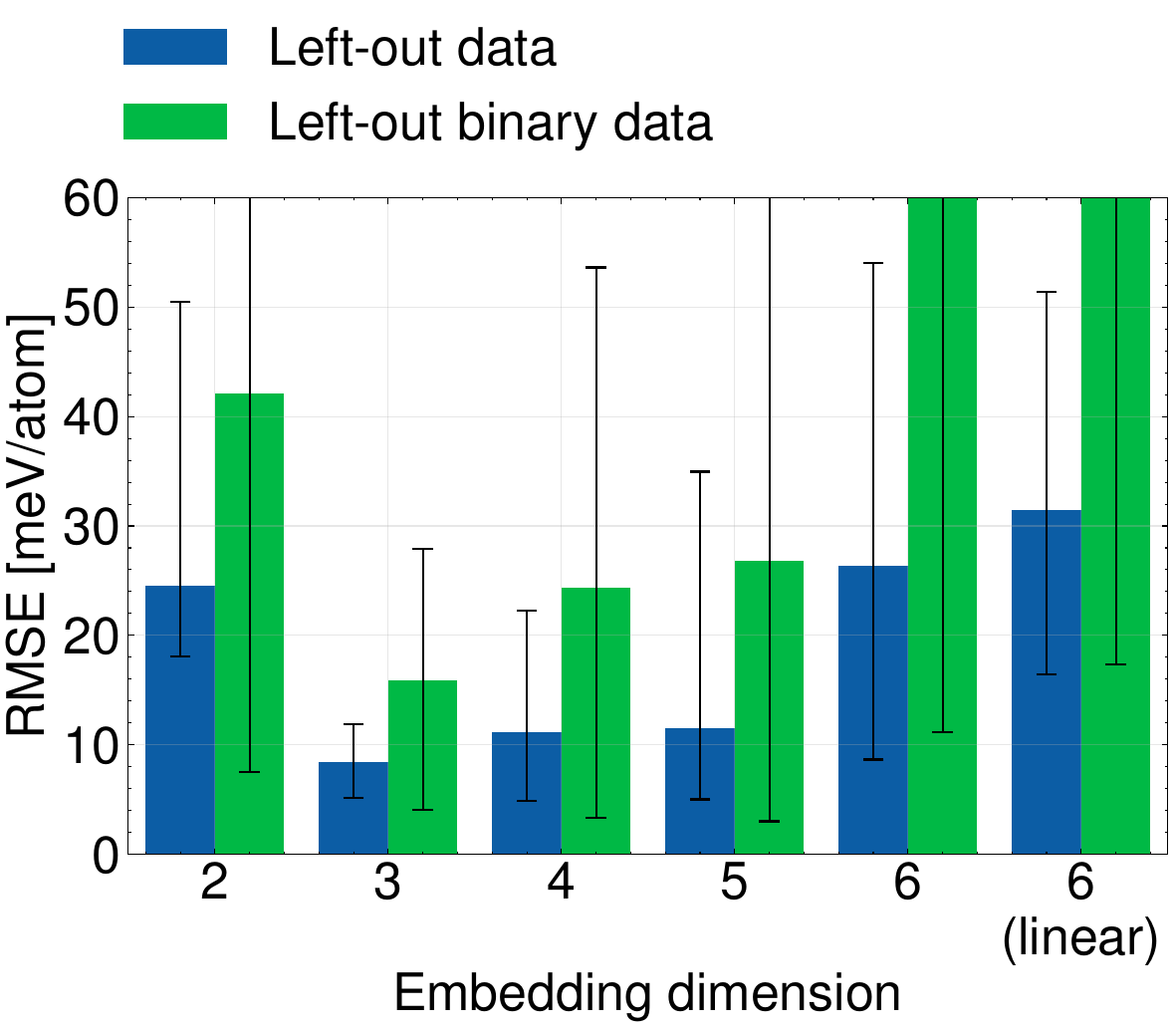}
	\caption{\textbf{Extrapolating the formation energies over left-out element pairs.} Average RMSE computed over all left-out element pairs for eCE models with embedding dimensionality from 2-6 and a linear CE model. Training sets are composed of all orderings except a specific pair of elements. Blue bars correspond to the errors computed over all left-out data. Green bars are the errors computed over only binary orderings of the left-out element pair. The spread for each bar is the minimum and maximum errors computed over all pairs of left-out elements.}
	\label{fig:binary_dropout}
\end{figure}

Training datasets for conventional CE models usually contain orderings that span the entire composition space of an alloy. As a result, multicomponent alloys with 3 or more chemical species can require orders of magnitude more data than simpler binary alloys. eCE models are able to leverage chemical similarities between alloying elements to extrapolate into unsampled regions of composition space. \Cref{fig:binary_dropout_example} compares two models that are trained on all configurations except those containing both molybdenum and tantalum. The linear CE and 3-eCE models should be unable to learn the interactions between Mo and Ta as configurations containing both elements are not included in the training dataset. Surprisingly, the 3-eCE model in \cref{fig:binary_dropout_example} reproduces the ordering energies of binary Mo-Ta configurations. 3-eCE is able to reproduce several ground states and the overall shape of the convex hull. In contrast, the conventional linear CE of \cref{fig:binary_dropout_example} fails to reproduce the energies of binary orderings. The failure of the conventional CE is not surprising as the model has to extrapolate into the binary Mo-Ta space. On the other hand, the 3-eCE model learns from chemical similarities in the dataset to effectively extrapolate into unseen composition spaces.

\Cref{fig:binary_dropout} benchmarks the extrapolation ability of eCE models across all possible pairs of left-out elements. Similar to \cref{fig:binary_dropout_example}, 15 separate training datasets were constructed by including all configurations from the senary dataset, except those containing a specific pair of elements. Each training dataset was then used to parameterize five eCE models with chemical embedding dimensions ranging from 2 to 6. A linear senary CE was also parameterized for each dataset. The linear CE serves as a baseline against which we compare the predictions of eCE. The validation errors computed over all left-out data and the left-out binary configurations are shown in \cref{fig:binary_dropout}. The benchmarks of \cref{fig:binary_dropout} are similar to the leave one out cross validation (LOOCV) metric. Rather than individual datapoints being left out of the dataset, entire alloys are used to cross-validate models.

\Cref{fig:binary_dropout} clearly demonstrates that eCE models are able to extrapolate into unseen composition spaces with significantly higher accuracy than conventional CE models.
Average extrapolation errors range from $\approx$30 meV/atom for a linear CE to $\approx8$ meV/atom for 3-eCE models. eCE models with 3 or 4 embedding dimensions are found to have the smallest extrapolation error, while all other eCE models have significantly larger errors. Although the average energy errors of 3-eCE models are small, we do find some degree of sensitivity to the exact pair of elements left out of the training dataset. \Cref{fig:binary_dropout} shows the range of energy errors across all 15 pairs of elements that are left out of the training dataset. For instance, the 3-eCE model is found to have extrapolation errors on left-out binary configurations that range from $\approx5 - 30$ meV/atom. The binary alloy comprised of the left-out elements corresponds to the most challenging datapoints to reproduce with eCE models. As a result the average prediction errors over configurations containing just the left-out elements (green bars in \cref{fig:binary_dropout}) is higher than the prediction error over configurations containing other elements in addition to the left-out pair. The highest extrapolation errors are found to occur for either alloys containing either Cr and V or Cr and Nb. This suggests that even with chemical compression, the bonding between some elements may be too complex to extrapolate from interactions in other alloys.

Embedded cluster expansions are able to utilize chemical similarities to achieve low prediction errors in chemical spaces that would be entirely extrapolative when using conventional CE. For instance, as shown in \cref{fig:binary_dropout}, eCE models with embedding dimensions of 3 and 4 perform significantly better than eCE models with higher embedding dimensions. As the number of effective chemical species increase, the eCE model treats chemical species with a greater degree of independence. Thus, higher dimensional embeddings lose the ability to leverage chemical trends to lower prediction errors. In \cref{fig:binary_dropout}, where all configurations containing a pair of elements are left out, 6-eCE models perform poorly as they are fully extrapolative. In contrast, 3-eCE models utilize chemical trends to achieve lower prediction errors.

\begin{figure}[h]
	\centering
	\includegraphics[width=0.48\textwidth]{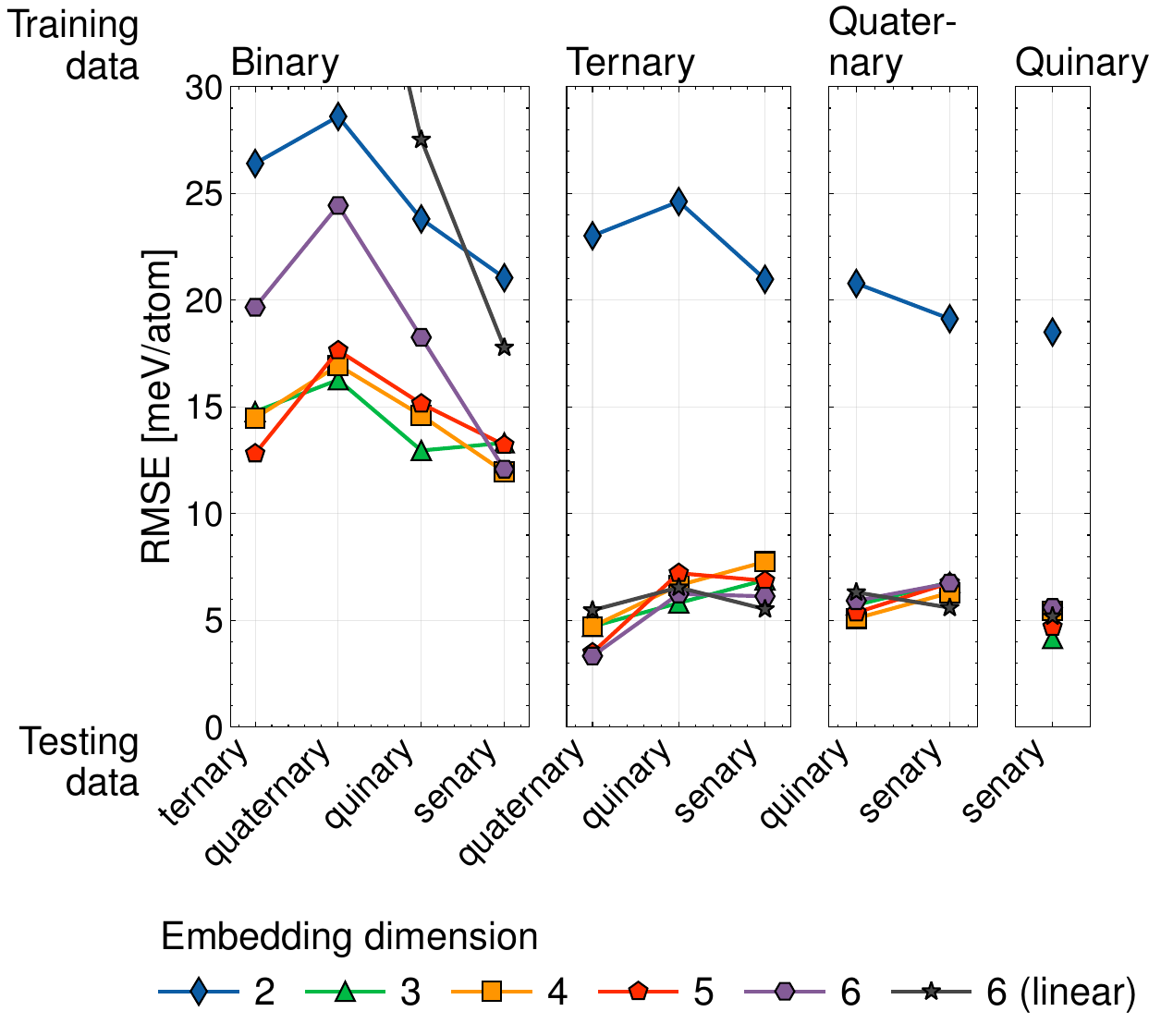}
	\caption{\textbf{Extrapolation error in multicomponent chemical space.} Comparison of the RMSE with the number of chemical elements in the validation dataset. Each panel in the figure corresponds to a training dataset constructed by choosing orderings with at most the number of elements indicated. For instance, the ternary training dataset included all ternary and binary orderings. The validation errors are then computed for higher order systems. eCE models with embedding dimensions ranging from 2-6 are compared with a linear model.}
	\label{fig:dataset_complexity}
\end{figure}

Similar to the CALPHAD method, CE models of multicomponent alloys can be parameterized starting from training data that spans lower-order constituent systems such as binaries, ternaries, etc \cite{goiri2018}. \Cref{fig:dataset_complexity} tests the ability of eCE models to extrapolate into multicomponent space when it is trained on lower-order chemical spaces. All models in the first panel of \cref{fig:dataset_complexity} were parameterized based on a dataset containing only binary orderings of the elements in groups 5 and 6 of the periodic table. The models were then tested on higher-order composition spaces containing 3 or more elements. As shown in \cref{fig:dataset_complexity}, 3-eCE models have extrapolation errors ranging between $\approx12$ meV/atom, with the more complex orderings being better determined than the simpler ternary orderings. In contrast, 2-eCE  and linear CE models have significantly larger errors. Similar to \cref{fig:binary_dropout_example,fig:binary_dropout,fig:hyperparameter_optimization}, 2-eCE lacks sufficient flexibility to capture the chemical interactions of this alloy system. The energies of binary datapoints are likely insufficient to accurately describe the multicomponent energetics in the senary alloy as 3-eCE models parameterized over separate random instantiations resulted in errors ranging from 10 to 20 meV/atom.

Adding configurations with orderings of up to 3 elements drastically lowers the extrapolation error of most CE models. In fact, adding configurations beyond ternary orderings shows only marginal improvement in extrapolation errors ($\approx3-4$ meV/atom). While configurations with multiple alloying elements may be critical to accurately capturing low-energy ground states, the results of \cref{fig:dataset_complexity} suggest that the senary alloy formed from elements in groups 5 and 6 of the periodic table are primarily composed of unary, binary and ternary interactions. This is in excellent agreement with our findings from previous sections indicating that 3-eCE models are able to accurately reproduce the ordering energetics of this senary alloy.

\subsection{Finite-temperature predictions}
\label{sec:finite-temp-pred}

\begin{figure}[h]
	\centering
	\includegraphics[width=0.47\textwidth]{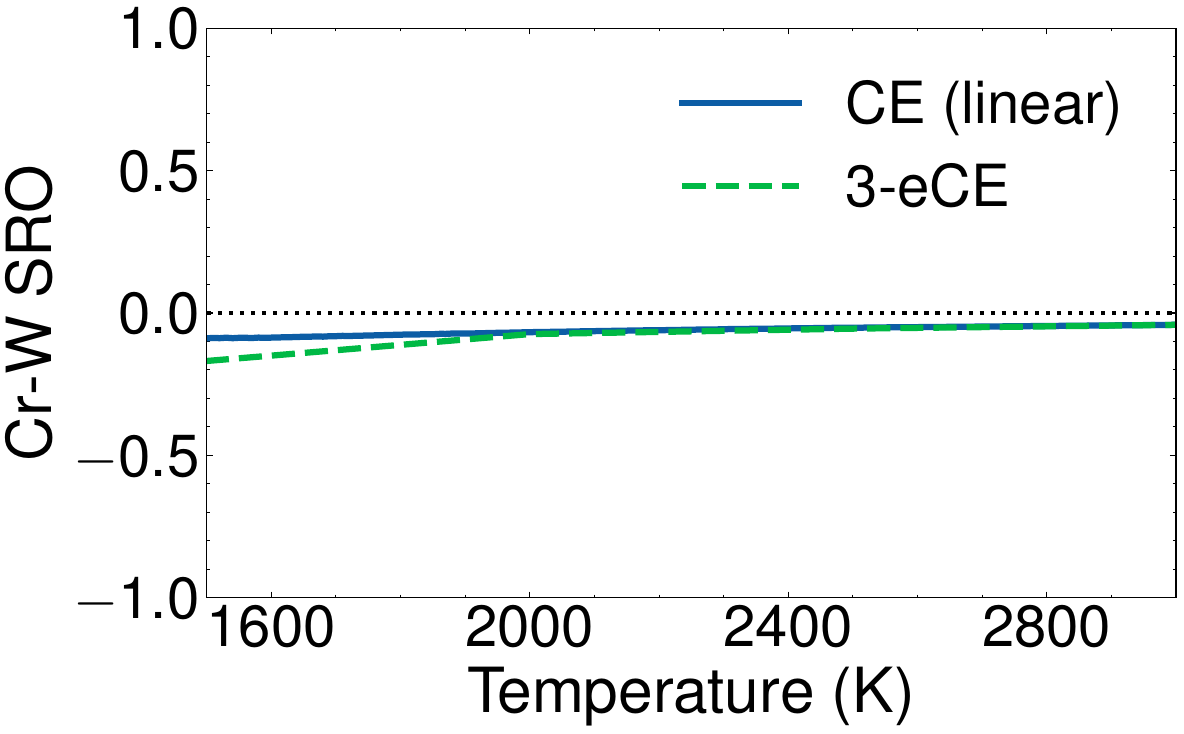}
	\caption{\textbf{Warren-Cowley short-range order (SRO) parameters for Cr-W pairs in an equiatomic Cr-W binary alloy.} The SRO values computed from canonical Monte-Carlo simulations with a conventional CE is compared against the values based on a 3-eCE model.}
	\label{fig:SRO_comparison_CrW}
\end{figure}

Surrogate models such as CE are ultimately used to compute finite-temperature quantities such as free energies, heat capacities, or short-range order parameters. \Cref{fig:SRO_comparison_CrW} compares the nearest neighbor Warren-Cowley short-range order parameters (SRO) in the Cr-W binary alloy computed with a linear CE to the values obtained from a 3-eCE model. The linear CE was parameterized with binary orderings of Cr and W in our dataset. Clusters containing up to 4 sites and a distance of 5.5$\text{\AA}$ were included in the model that achieved an RMSE of 3.5 meV/atom. The 3-eCE model is identical to the model used in \cref{fig:convex_hull_compare}. Canonical Monte-Carlo simulations are employed to compute ensemble averages of the Cr-W SRO as a function of temperature. Both models show essentially identical values of the SRO at elevated temperatures and start to slightly deviate at temperatures approaching $\approx 1600$K. The Cr-W SRO value is found to be slightly negative, indicating an elevated number of Cr-W pairs as compared to the truly disordered alloy. Similar values for the Cr-W alloy have been computed recently with both CE and off-lattice interatomic potentials\cite{smith2024}.  The agreement in finite-temperature properties indicates that the 3-eCE model has comparable accuracies to a conventional CE.

\begin{figure}[h]
	\centering
	\includegraphics[width=0.47\textwidth]{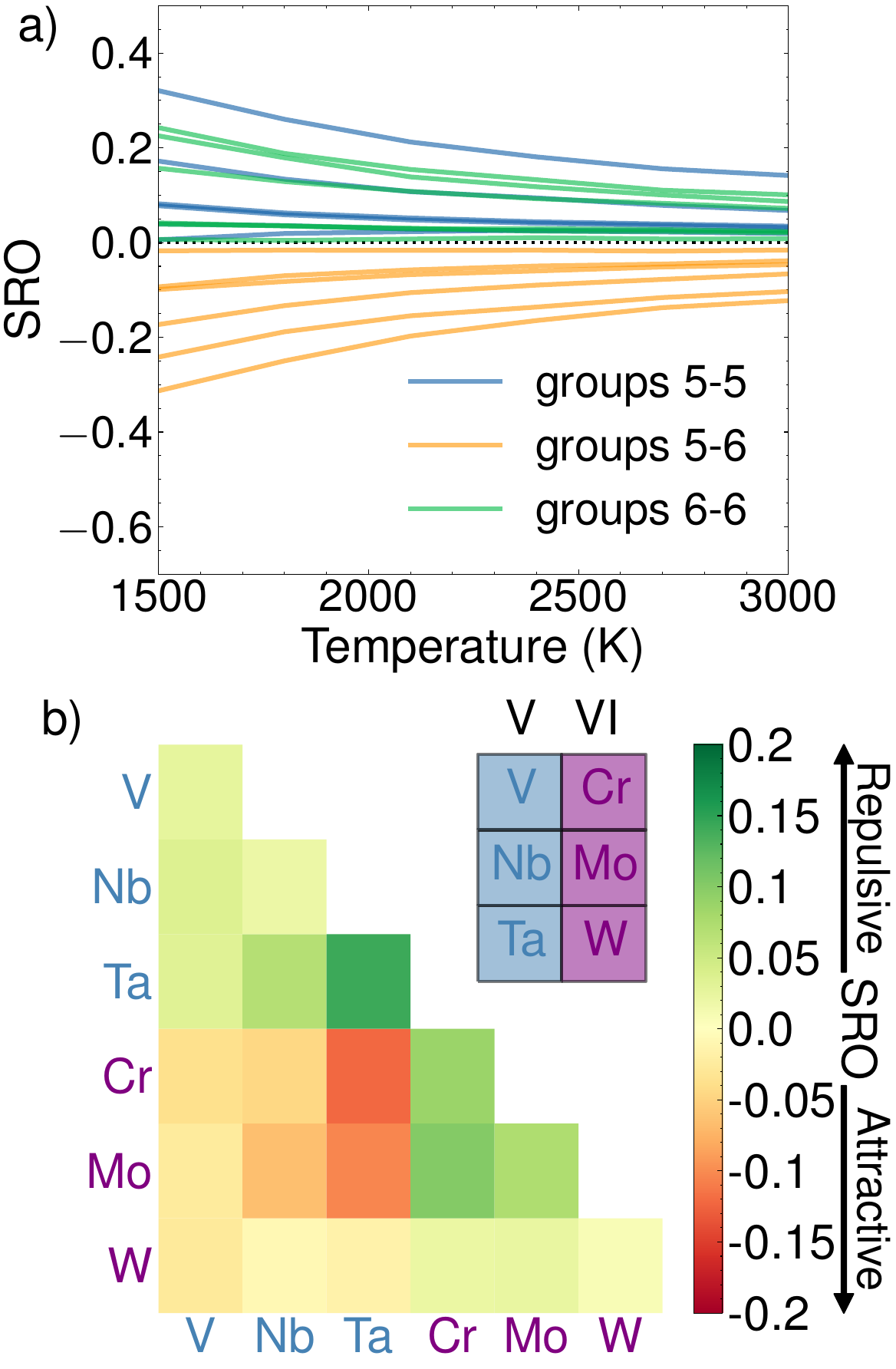}
	\caption{\textbf{Warren-Cowley short-range order (SRO) parameters in an equiatomic senary V-Nb-Ta-Cr-Mo-W alloy.} The SRO is computed with a 3-eCE model and canonical Monte-Carlo simulations. (a) SRO variation with temperature for pairs of elements labeled by the group number, i.e., group 5 elements with group 5 elements in blue, group 5 elements with group 6 elements in orange, and group 6 elements with group 6 elements in green. (b) SRO values for all pairs of elements at 3000K.}
	\label{fig:SRO_senary}
\end{figure}

Having established the accuracy of eCE, we employ the 3-eCE model to compute finite-temperature SRO values in an equiatomic senary V-Nb-Ta-Cr-Mo-W alloy (\cref{fig:SRO_senary}). \Cref{fig:SRO_senary}a shows the SRO values for elements plotted based on their group numbers. Pairs of elements in the same group display positive values of SRO, corresponding to fewer nearest neighbor pairs as compared with a random alloy. Choosing one element from group 5 and another from group 6 results in negative values of SRO. The short-range order parameter values for all pairs of elements at 3000 K are shown in \cref{fig:SRO_senary}b. Element pairs involving one element from group 5 and another from group 6 either have SRO values close to zero or are strongly negative even at elevated temperatures. Our finite-temperature simulations suggest that pairs of elements from groups 5 and 6 are attractive and there should be an elevated number of such pairs even at elevated temperatures. In turn, this causes a decrease in the number of element pairs from the same group.

\section{Discussion}
\label{sec:discussion}

Cluster expansions are the tool of choice to study phase transformations and finite temperature properties of multicomponent alloys. The embedded cluster expansion (eCE) model introduced in this study leverages chemical similarities between elements to construct CE type models for alloys containing several elements. The eCE model simultaneously learns a lower-dimensional embedding of site basis functions along with the regression coefficients of a site-centric energy model. The site energies within eCE use cluster functions constructed from the transformed site functions that lie in a lower-dimensional space. As fewer site functions are required to describe occupants at any given site, there is a drastic reduction in the number of cluster functions. The results of \cref{fig:hyperparameter_optimization} show that eCE models can reach accuracies comparable to conventional CE models. Zero Kelvin phase stability predicted by eCE models is also found to be quantitatively accurate (\cref{fig:convex_hull_compare}). Allowing the model to learn chemical similarities between elements enables robust extrapolation into unsampled chemical spaces. Despite leaving pairs of elements out in \cref{fig:binary_dropout_example,fig:binary_dropout}, eCE models capture the energetics of the left-out alloy. The results of \cref{fig:dataset_complexity} suggest that orderings from simpler chemical sub-systems may be sufficient to capture multicomponent interactions in concentrated alloys. Finite temperature properties such as short-range order are also found to be well-reproduced by eCE models (\cref{fig:SRO_comparison_CrW}), allowing us to investigate trends in SRO for the multicomponent senary alloy (\cref{fig:SRO_senary}).

\begin{figure}[h]
	\centering
	\includegraphics[width=0.47\textwidth]{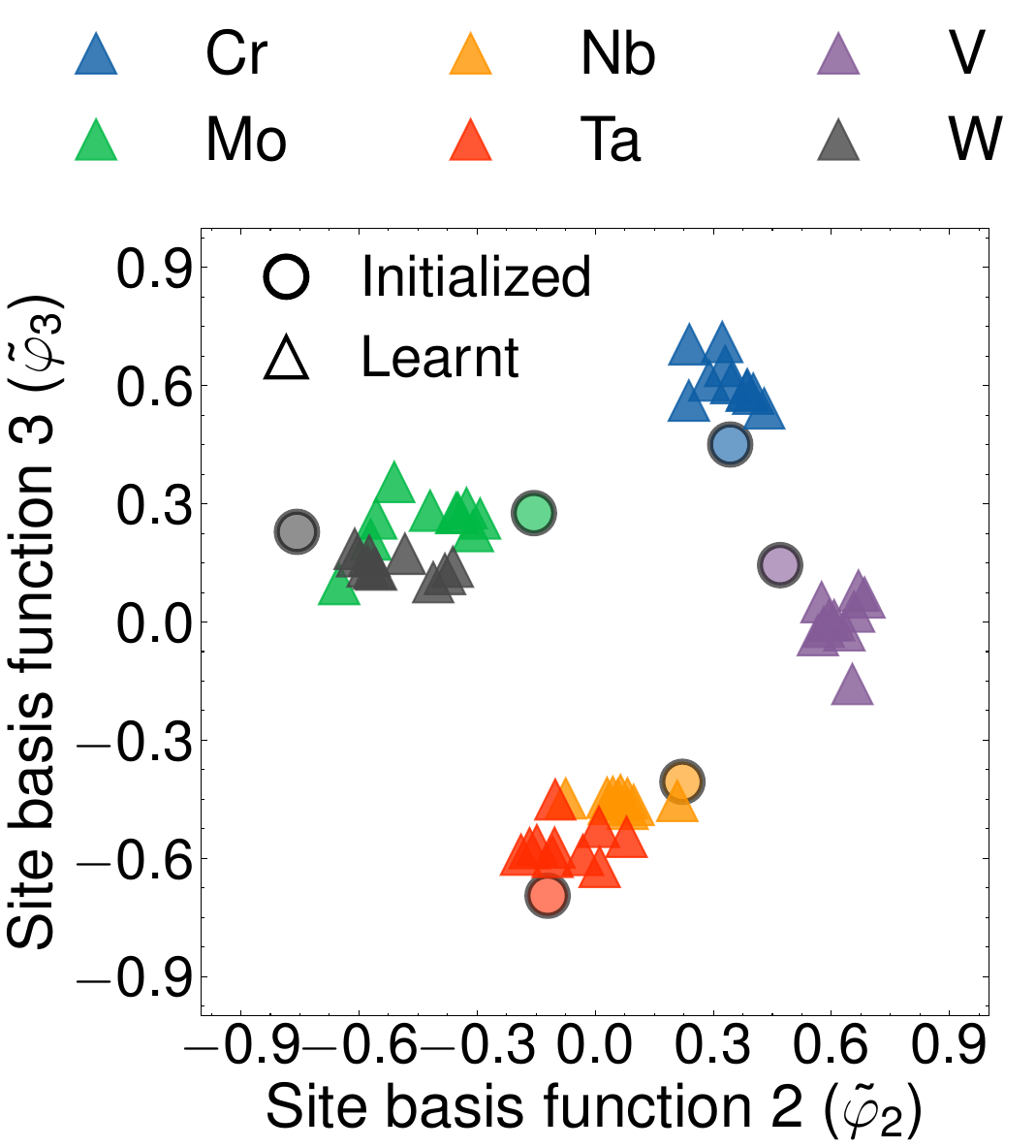}
	\caption{\textbf{Site basis function values learnt by 3-eCE models.} The initial values of the site basis functions are shown as circles. The final value learnt by the 3-eCE model over 10 separate models is shown as triangles. Each element is denoted with a different color.}
	\label{fig:projections}
\end{figure}

The values of the site basis functions learnt by an eCE model can shed light on chemical similarities between alloying elements. Site function values learned by a 3-eCE model for the six refractory elements in groups 5 and 6 of the periodic table are shown in \cref{fig:projections}. 10 separate 3-eCE models were parameterized starting from an identical initialization for the elements of the embedding matrix ($\mathcal{T}$ in \cref{eq:site_basis_function_projection}). The initial values of the site basis functions for each element are shown as circles and the final values learnt by the 3-eCE model are represented as triangles in the figure. 

\Cref{fig:projections} shows several chemical trends across all 3-eCE models. Pairs of elements, such as molybdenum and tungsten or tantalum and niobium have similar site basis function values. Chromium and vanadium on the other hand are separated from the site basis function values of the other elements. Both elements are not found to cluster with any other element in \cref{fig:projections}. The clustering of site basis function values can be correlated with chemical similarities between elements. For example, molybdenum and tungsten have similar metallic radii and belong to group 6 of the periodic table. This results in very similar chemical interactions as reflected by embedded values of the site basis functions for both elements in \cref{fig:projections}.
In contrast, the other element of group 6, chromium, is smaller than Mo and W. This results in qualitatively different interactions of Cr and causes the model to separate the element in the projected space.
The elements of group 5 in the periodic table have very different site basis function values than the elements of group 6. The chemically similar elements, niobium and tantalum, have embedded site function values that are in close proximity, while vanadium, that is smaller than both elements is clearly differentiated in \cref{fig:projections}.
The lack of elements similar to Cr and V may be related to the large extrapolation errors observed for Cr-V containing configurations in \cref{fig:binary_dropout}. Some degree of scatter is evident over the models due to the stochasticity of the gradient descent technique used to minimize \cref{eq:loss_function_training}. Nevertheless, all parameterizations show similar trends.  

\begin{figure}[h]
	\centering
	\includegraphics[width=0.47\textwidth]{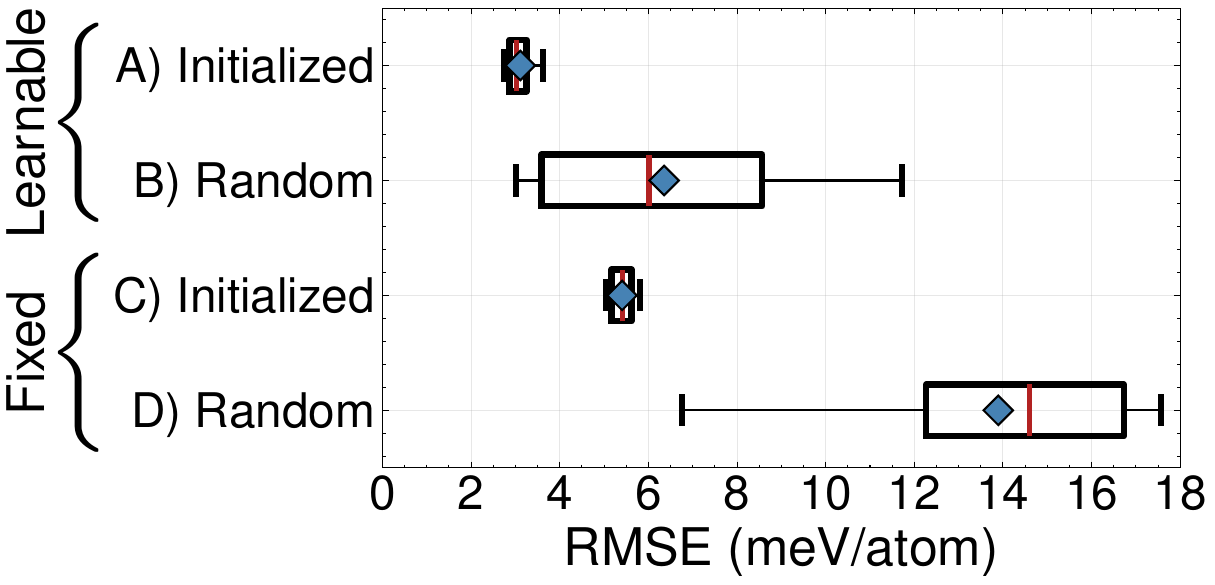}
	\caption{\textbf{Effect of the embedding matrix initialization and learning scheme on validation error.} Box plots of the validation RMSE computed for 10 different 3-eCE parameterizations for four different projection schemes. Each group of models is either initialized with an embedding matrix as described in \cref{sec:methods} or with a random embedding matrix. One set of models are allowed to learn the embedding matrix starting from the initial value, while in the other set of models the embedding matrix is fixed.}
	\label{fig:projection_initialization}
\end{figure}

Our results also suggest that learning the transformation matrix, $\mathcal{T}$, and careful initialization of the embedding matrix is crucial to obtaining predictive models. \Cref{fig:projection_initialization} compares the validation errors of four different learning schemes. Two sets of models are parameterized while allowing for the transformation matrix to be learnt during model training. In the other two groups of models, the embedding matrix is fixed to its initial value. We also attempt two different initialization strategies. The first group of models are initialized based on similarities in the chemical properties of each element (details are outlined in \cref{sec:methods}). Random orthogonal projection vectors are used for the initial embedding matrix in the other models. \Cref{fig:projection_initialization} shows the range of validation errors obtained over 10 instantiations of a 3-eCE model. Random initializations result in a large variance of the validation error. Difficulty in learning predictive eCE models from random initializations could be due to the existence of multiple local minima of the loss function. All models that are initialized with a transformation matrix containing some chemical information are able to achieve significantly lower prediction errors than random embedding matrices. The 3-eCE model with the lowest validation error that was initialized with a random projection matrix was also able to learn elemental similiarities like those shown in \cref{fig:projections}. Interestingly, a learnable transformation matrix seems to be necessary to enhance the predictive power of the model. Comparing the models with chemically informed initializations of the transformation matrix, \cref{fig:projection_initialization} suggests that learning the best embedding matrix could lower errors by $\approx 3$ meV/atom. This can also be seen in the reduced spread of validation errors for random initializations with a learnable embedding matrix. While the benefits of learning the embedding matrix for the senary refractory alloy system are not very large, this may be important for multicomponent alloys with more alloying elements.

The eCE framework of \cref{eq:site_basis_function_projection,eq:symmetrized_projected_cluster_function,eq:site_centric_projected_energies}, is similar to recently proposed chemical embedding schemes for off-lattice interatomic potentials\cite{willatt2018,artrith2017,gastegger2018,huo2022a,darby2022}. The application of chemical compression schemes to off-lattice models have enabled researchers to investigate alloys containing several elements\cite{mazitov2024,lopanitsyna2023}. Parameterizing such off-lattice models can be very expensive, often requiring tens of thousands of calculations. Additionally, as suggested in a recent study\cite{mazitov2024}, resolving the small energy differences between competing intermetallic orderings with a general interatomic potential can be challenging. The eCE surrogate model could provide sufficient energy accuracy and computational speed to bridge this gap and allow researchers to study complex alloy thermodynamics. As shown by \cref{fig:hyperparameter_optimization}, relatively small training datasets are required to parameterize on-lattice models with chemical compression schemes. The eCE models are sufficiently flexible to extrapolate into higher dimensional composition spaces. Additionally, computing finite-temperature properties from eCE models is relatively straight forward and computationally cheap. This could enable alloy designers to rapidly screen materials for compositions with desirable properties through eCE based surrogate models. Promising alloy chemistries that require more accurate simulations that account for all sources of entropy can subsequently be studied with bespoke interatomic potentials. 

The chemical flexibility of the eCE formalism and the smaller dataset sizes needed to parameterize these models will enable the systematic exploration of high-dimensional composition spaces. eCE models will provide significant advantages against conventional CE in alloys where some chemical trends (similarities or dissimilarities) exist between groups of elements. In materials where elements are chemically uncorrelated or with very complex chemical trends, eCE models may require larger embedding dimensions, perhaps even approaching the number of elements in the alloy. Benchmarks such as the learning curves of \cref{fig:hyperparameter_optimization} can be used to discern the appropriate dimensionality of the embedding space. The eCE formalism is also subject to several of the same restrictions as conventional CE. For instance, alloys with long-range interactions, or significant structural relaxations will continue to remain challenging to parameterize with eCE models. The problem can be somewhat alleviated through structure matching algorithms\cite{thomas2021} to prune large relaxations out of training databases and the explicit inclusion of additional terms accounting for long-range interactions\cite{kitchaev2018,laks1992}. Further, all significant sources of entropy will need to be included in the formalism to enable a rigorous comparison with experiment. This may require the coupling of eCE models to other site degrees of freedom such as magnetic moments, vibrations or lattice distortions. The extension of the eCE model to such coupled effective Hamiltonians can be done similarly to existing methods that couple site occupancy with other discrete or continuous degrees of freedom.

\section{Methods}
\label{sec:methods}

\subsection{DFT calculations}
Formation energies of 4083 symmetrically distinct orderings between elements of groups 5 and 6 (Cr-Mo-Nb-Ta-V-W) are calculated with the generalized gradient approximation (GGA-PBE) to density functional theory (DFT) and projector augmented-wave (PAW) pseudopotentials as implemented in the \emph{Vienna Ab-Initio Simulation Package (VASP)} \cite{blochl1994,kresse1993,kresse1996,kresse1996a}. A plane-wave cutoff energy of 550 eV with a k-point grid density of 55\AA$ $ and smearing of 0.1 eV were used to relax the positions of atoms and lattice parameters of all orderings. Symmetrically distinct orderings on a parent bcc crystal structure are enumerated with the \texttt{CASM} code\cite{vanderven2018}. 2487 symmetrically distinct orderings on supercells of sizes up to 12 are enumerated within the binary and ternary sub-systems of the senary V-Nb-Ta-Cr-Mo-W alloy. Symmetrically distinct equiatomic orderings in the quaternary, quinary, and senary alloys are enumerated in supercells containing up to 6 atoms. 387 random arrangements of the 6 elements in a supercell containing 8 atoms are also included in the training dataset. SRO values are computed with canonical Monte Carlo simulations performed in a 10x10x10 supercell of the conventional bcc cell. The short-range order parameters are computed by averaging over 1000 Monte-Carlo passes.

\subsection{Embedded Cluster Expansion}
The embedded cluster expansion was implemented in \texttt{python} using the \texttt{PyTorch} library. Gradient descent of the loss function in \cref{eq:loss_function_training} is performed with the stochastic Adam algorithm for 100 epochs. A learning rate scheduler is applied and overfitting is controlled through the L2-regularization. A graph for the site energy is built within \texttt{PyTorch} starting from Chebyshev site basis functions that are projected into a lower dimensional space through a learnable embedding matrix. Symmetrized site-centric cluster functions constructed as tensor products of the embedded site basis functions are used as input to a 4-layer neural network ($32\times32\times8\times1$) that uses the ReLU activation function for each node except the final layer, where we use a linear activation function. The rows of the learnable linear transformation $\mathcal{T}$ are re-normalized after each iteration.

eCE models were initialized with a projection matrix, $\mathcal{T}$ computed from chemical properties of each element. 8 elemental properties (atomic number, radius, electronegativity, density, melting point, bulk modulus, Youngs modulus and Brinell hardness) were collected for each element from \texttt{pymatgen}\cite{ong2013}. The material properties for each element were used to form the columns of a matrix, $A$. The rows of $A$ were standardized to have zero mean and a standard deviation of 1. An embedding matrix, with an embedding dimensionality of $k$ was initialized with the first $k$ right-singular vectors of $A$.

\section{Acknowledgements}
This research was partially supported by the NCCR MARVEL, a National Centre of Competence in Research, funded by the Swiss National Science Foundation (grant number 205602). Funding from the SNSF through project number 215178 is also acknowledged. We thank Prof. Michele Ceriotti for the helpful discussions.

\bibliography{references}

\begin{thebibliography}{67}%
\makeatletter
\providecommand \@ifxundefined [1]{%
 \@ifx{#1\undefined}
}%
\providecommand \@ifnum [1]{%
 \ifnum #1\expandafter \@firstoftwo
 \else \expandafter \@secondoftwo
 \fi
}%
\providecommand \@ifx [1]{%
 \ifx #1\expandafter \@firstoftwo
 \else \expandafter \@secondoftwo
 \fi
}%
\providecommand \natexlab [1]{#1}%
\providecommand \enquote  [1]{``#1''}%
\providecommand \bibnamefont  [1]{#1}%
\providecommand \bibfnamefont [1]{#1}%
\providecommand \citenamefont [1]{#1}%
\providecommand \href@noop [0]{\@secondoftwo}%
\providecommand \href [0]{\begingroup \@sanitize@url \@href}%
\providecommand \@href[1]{\@@startlink{#1}\@@href}%
\providecommand \@@href[1]{\endgroup#1\@@endlink}%
\providecommand \@sanitize@url [0]{\catcode `\\12\catcode `\$12\catcode
  `\&12\catcode `\#12\catcode `\^12\catcode `\_12\catcode `\%12\relax}%
\providecommand \@@startlink[1]{}%
\providecommand \@@endlink[0]{}%
\providecommand \url  [0]{\begingroup\@sanitize@url \@url }%
\providecommand \@url [1]{\endgroup\@href {#1}{\urlprefix }}%
\providecommand \urlprefix  [0]{URL }%
\providecommand \Eprint [0]{\href }%
\providecommand \doibase [0]{https://doi.org/}%
\providecommand \selectlanguage [0]{\@gobble}%
\providecommand \bibinfo  [0]{\@secondoftwo}%
\providecommand \bibfield  [0]{\@secondoftwo}%
\providecommand \translation [1]{[#1]}%
\providecommand \BibitemOpen [0]{}%
\providecommand \bibitemStop [0]{}%
\providecommand \bibitemNoStop [0]{.\EOS\space}%
\providecommand \EOS [0]{\spacefactor3000\relax}%
\providecommand \BibitemShut  [1]{\csname bibitem#1\endcsname}%
\let\auto@bib@innerbib\@empty
\bibitem [{\citenamefont {Sanchez}\ \emph {et~al.}(1984)\citenamefont
  {Sanchez}, \citenamefont {Ducastelle},\ and\ \citenamefont
  {Gratias}}]{sanchez1984}%
  \BibitemOpen
  \bibfield  {author} {\bibinfo {author} {\bibfnamefont {J.}~\bibnamefont
  {Sanchez}}, \bibinfo {author} {\bibfnamefont {F.}~\bibnamefont
  {Ducastelle}},\ and\ \bibinfo {author} {\bibfnamefont {D.}~\bibnamefont
  {Gratias}},\ }\bibfield  {title} {\bibinfo {title} {Generalized cluster
  description of multicomponent systems},\ }\href
  {https://doi.org/10.1016/0378-4371(84)90096-7} {\bibfield  {journal}
  {\bibinfo  {journal} {Physica A: Statistical Mechanics and its Applications}\
  }\textbf {\bibinfo {volume} {128}},\ \bibinfo {pages} {334} (\bibinfo {year}
  {1984})}\BibitemShut {NoStop}%
\bibitem [{\citenamefont {Fontaine}(1994)}]{fontaine1994}%
  \BibitemOpen
  \bibfield  {author} {\bibinfo {author} {\bibfnamefont {D.~D.}\ \bibnamefont
  {Fontaine}},\ }\bibfield  {title} {\bibinfo {title} {Cluster {{Approach}} to
  {{Order-Disorder Transformations}} in {{Alloys}}},\ }in\ \href
  {https://doi.org/10.1016/S0081-1947(08)60639-6} {\emph {\bibinfo {booktitle}
  {Solid {{State Physics}}}}},\ Vol.~\bibinfo {volume} {47}\ (\bibinfo
  {publisher} {Elsevier},\ \bibinfo {year} {1994})\ pp.\ \bibinfo {pages}
  {33--176}\BibitemShut {NoStop}%
\bibitem [{\citenamefont {M{\"u}ller}\ and\ \citenamefont
  {Natarajan}(2024)}]{muller2024}%
  \BibitemOpen
  \bibfield  {author} {\bibinfo {author} {\bibfnamefont {Y.~L.}\ \bibnamefont
  {M{\"u}ller}}\ and\ \bibinfo {author} {\bibfnamefont {A.~R.}\ \bibnamefont
  {Natarajan}},\ }\bibfield  {title} {\bibinfo {title} {First-principles
  thermodynamics of precipitation in aluminum-containing refractory alloys},\
  }\href {https://doi.org/10.1016/j.actamat.2024.119995} {\bibfield  {journal}
  {\bibinfo  {journal} {Acta Materialia}\ }\textbf {\bibinfo {volume} {274}},\
  \bibinfo {pages} {119995} (\bibinfo {year} {2024})}\BibitemShut {NoStop}%
\bibitem [{\citenamefont {{Van der Ven}}\ \emph {et~al.}(2018)\citenamefont
  {{Van der Ven}}, \citenamefont {Thomas}, \citenamefont {Puchala},\ and\
  \citenamefont {Natarajan}}]{vanderven2018}%
  \BibitemOpen
  \bibfield  {author} {\bibinfo {author} {\bibfnamefont {A.}~\bibnamefont {{Van
  der Ven}}}, \bibinfo {author} {\bibfnamefont {J.}~\bibnamefont {Thomas}},
  \bibinfo {author} {\bibfnamefont {B.}~\bibnamefont {Puchala}},\ and\ \bibinfo
  {author} {\bibfnamefont {A.}~\bibnamefont {Natarajan}},\ }\bibfield  {title}
  {\bibinfo {title} {First-{{Principles Statistical Mechanics}} of
  {{Multicomponent Crystals}}},\ }\href
  {https://doi.org/10.1146/annurev-matsci-070317-124443} {\bibfield  {journal}
  {\bibinfo  {journal} {Annual Review of Materials Research}\ }\textbf
  {\bibinfo {volume} {48}},\ \bibinfo {pages} {27} (\bibinfo {year}
  {2018})}\BibitemShut {NoStop}%
\bibitem [{\citenamefont {Natarajan}\ and\ \citenamefont {{Van der
  Ven}}(2017{\natexlab{a}})}]{natarajan2017a}%
  \BibitemOpen
  \bibfield  {author} {\bibinfo {author} {\bibfnamefont {A.~R.}\ \bibnamefont
  {Natarajan}}\ and\ \bibinfo {author} {\bibfnamefont {A.}~\bibnamefont {{Van
  der Ven}}},\ }\bibfield  {title} {\bibinfo {title} {First-principles
  investigation of phase stability in the {{Mg-Sc}} binary alloy},\ }\href
  {https://doi.org/10.1103/PhysRevB.95.214107} {\bibfield  {journal} {\bibinfo
  {journal} {Physical Review B}\ }\textbf {\bibinfo {volume} {95}},\ \bibinfo
  {pages} {214107} (\bibinfo {year} {2017}{\natexlab{a}})}\BibitemShut
  {NoStop}%
\bibitem [{\citenamefont {Linder{\"a}lv}\ \emph {et~al.}(2022)\citenamefont
  {Linder{\"a}lv}, \citenamefont {Rahm},\ and\ \citenamefont
  {Erhart}}]{linderalv2022}%
  \BibitemOpen
  \bibfield  {author} {\bibinfo {author} {\bibfnamefont {C.}~\bibnamefont
  {Linder{\"a}lv}}, \bibinfo {author} {\bibfnamefont {J.~M.}\ \bibnamefont
  {Rahm}},\ and\ \bibinfo {author} {\bibfnamefont {P.}~\bibnamefont {Erhart}},\
  }\bibfield  {title} {\bibinfo {title} {High-{{Throughput Characterization}}
  of {{Transition Metal Dichalcogenide Alloys}}: {{Thermodynamic Stability}}
  and {{Electronic Band Alignment}}},\ }\href
  {https://doi.org/10.1021/acs.chemmater.2c01176} {\bibfield  {journal}
  {\bibinfo  {journal} {Chemistry of Materials}\ ,\ \bibinfo {pages}
  {acs.chemmater.2c01176}} (\bibinfo {year} {2022})}\BibitemShut {NoStop}%
\bibitem [{\citenamefont {Thomas}\ and\ \citenamefont {der
  Ven}(2013)}]{thomas2013}%
  \BibitemOpen
  \bibfield  {author} {\bibinfo {author} {\bibfnamefont {J.~C.}\ \bibnamefont
  {Thomas}}\ and\ \bibinfo {author} {\bibfnamefont {A.~V.}\ \bibnamefont {der
  Ven}},\ }\bibfield  {title} {\bibinfo {title} {Finite-temperature properties
  of strongly anharmonic and mechanically unstable crystal phases from first
  principles},\ }\href {https://doi.org/10.1103/PhysRevB.88.214111} {\bibfield
  {journal} {\bibinfo  {journal} {Physical Review B}\ }\textbf {\bibinfo
  {volume} {88}},\ \bibinfo {pages} {214111} (\bibinfo {year}
  {2013})}\BibitemShut {NoStop}%
\bibitem [{\citenamefont {Thomas}\ and\ \citenamefont {{Van der
  Ven}}(2014)}]{thomas2014}%
  \BibitemOpen
  \bibfield  {author} {\bibinfo {author} {\bibfnamefont {J.~C.}\ \bibnamefont
  {Thomas}}\ and\ \bibinfo {author} {\bibfnamefont {A.}~\bibnamefont {{Van der
  Ven}}},\ }\bibfield  {title} {\bibinfo {title} {Elastic properties and
  stress-temperature phase diagrams of high-temperature phases with
  low-temperature lattice instabilities},\ }\href
  {https://doi.org/10.1103/PhysRevB.90.224105} {\bibfield  {journal} {\bibinfo
  {journal} {Physical Review B}\ }\textbf {\bibinfo {volume} {90}},\ \bibinfo
  {pages} {224105} (\bibinfo {year} {2014})}\BibitemShut {NoStop}%
\bibitem [{\citenamefont {{van de Walle}}\ and\ \citenamefont
  {Ceder}(2002)}]{vandewalle2002a}%
  \BibitemOpen
  \bibfield  {author} {\bibinfo {author} {\bibfnamefont {A.}~\bibnamefont {{van
  de Walle}}}\ and\ \bibinfo {author} {\bibfnamefont {G.}~\bibnamefont
  {Ceder}},\ }\bibfield  {title} {\bibinfo {title} {The effect of lattice
  vibrations on substitutional alloy thermodynamics},\ }\href@noop {}
  {\bibfield  {journal} {\bibinfo  {journal} {Rev. Mod. Phys.}\ }\textbf
  {\bibinfo {volume} {74}},\ \bibinfo {pages} {35} (\bibinfo {year}
  {2002})}\BibitemShut {NoStop}%
\bibitem [{\citenamefont {Kadkhodaei}\ \emph {et~al.}(2017)\citenamefont
  {Kadkhodaei}, \citenamefont {Hong},\ and\ \citenamefont {Van
  De~Walle}}]{kadkhodaei2017}%
  \BibitemOpen
  \bibfield  {author} {\bibinfo {author} {\bibfnamefont {S.}~\bibnamefont
  {Kadkhodaei}}, \bibinfo {author} {\bibfnamefont {Q.-J.}\ \bibnamefont
  {Hong}},\ and\ \bibinfo {author} {\bibfnamefont {A.}~\bibnamefont {Van
  De~Walle}},\ }\bibfield  {title} {\bibinfo {title} {Free energy calculation
  of mechanically unstable but dynamically stabilized bcc titanium},\ }\href
  {https://doi.org/10.1103/PhysRevB.95.064101} {\bibfield  {journal} {\bibinfo
  {journal} {Physical Review B}\ }\textbf {\bibinfo {volume} {95}},\ \bibinfo
  {pages} {064101} (\bibinfo {year} {2017})}\BibitemShut {NoStop}%
\bibitem [{\citenamefont {Kitchaev}\ and\ \citenamefont {Van
  Der~Ven}(2021)}]{kitchaev2021a}%
  \BibitemOpen
  \bibfield  {author} {\bibinfo {author} {\bibfnamefont {D.~A.}\ \bibnamefont
  {Kitchaev}}\ and\ \bibinfo {author} {\bibfnamefont {A.}~\bibnamefont {Van
  Der~Ven}},\ }\bibfield  {title} {\bibinfo {title} {Tuning magnetic
  antiskyrmion stability in tetragonal inverse {{Heusler}} alloys},\ }\href
  {https://doi.org/10.1103/PhysRevMaterials.5.124408} {\bibfield  {journal}
  {\bibinfo  {journal} {Physical Review Materials}\ }\textbf {\bibinfo {volume}
  {5}},\ \bibinfo {pages} {124408} (\bibinfo {year} {2021})}\BibitemShut
  {NoStop}%
\bibitem [{\citenamefont {Decolvenaere}\ \emph {et~al.}(2017)\citenamefont
  {Decolvenaere}, \citenamefont {Gordon}, \citenamefont {Seshadri},\ and\
  \citenamefont {Van Der~Ven}}]{decolvenaere2017}%
  \BibitemOpen
  \bibfield  {author} {\bibinfo {author} {\bibfnamefont {E.}~\bibnamefont
  {Decolvenaere}}, \bibinfo {author} {\bibfnamefont {M.}~\bibnamefont
  {Gordon}}, \bibinfo {author} {\bibfnamefont {R.}~\bibnamefont {Seshadri}},\
  and\ \bibinfo {author} {\bibfnamefont {A.}~\bibnamefont {Van Der~Ven}},\
  }\bibfield  {title} {\bibinfo {title} {First-principles investigation of
  competing magnetic interactions in ( {{Mn}} , {{Fe}} ) {{Ru}} 2 {{Sn
  Heusler}} solid solutions},\ }\href
  {https://doi.org/10.1103/PhysRevB.96.165109} {\bibfield  {journal} {\bibinfo
  {journal} {Physical Review B}\ }\textbf {\bibinfo {volume} {96}},\ \bibinfo
  {pages} {165109} (\bibinfo {year} {2017})}\BibitemShut {NoStop}%
\bibitem [{\citenamefont {Decolvenaere}\ \emph {et~al.}(2019)\citenamefont
  {Decolvenaere}, \citenamefont {Levin}, \citenamefont {Seshadri},\ and\
  \citenamefont {Van Der~Ven}}]{decolvenaere2019}%
  \BibitemOpen
  \bibfield  {author} {\bibinfo {author} {\bibfnamefont {E.}~\bibnamefont
  {Decolvenaere}}, \bibinfo {author} {\bibfnamefont {E.}~\bibnamefont {Levin}},
  \bibinfo {author} {\bibfnamefont {R.}~\bibnamefont {Seshadri}},\ and\
  \bibinfo {author} {\bibfnamefont {A.}~\bibnamefont {Van Der~Ven}},\
  }\bibfield  {title} {\bibinfo {title} {Modeling magnetic evolution and
  exchange hardening in disordered magnets: {{The}} example of {{Mn}} 1 - x
  {{Fe}} x {{Ru}} 2 {{Sn Heusler}} alloys},\ }\href
  {https://doi.org/10.1103/PhysRevMaterials.3.104411} {\bibfield  {journal}
  {\bibinfo  {journal} {Physical Review Materials}\ }\textbf {\bibinfo {volume}
  {3}},\ \bibinfo {pages} {104411} (\bibinfo {year} {2019})}\BibitemShut
  {NoStop}%
\bibitem [{\citenamefont {Drautz}\ and\ \citenamefont
  {F{\"a}hnle}(2004)}]{drautz2004a}%
  \BibitemOpen
  \bibfield  {author} {\bibinfo {author} {\bibfnamefont {R.}~\bibnamefont
  {Drautz}}\ and\ \bibinfo {author} {\bibfnamefont {M.}~\bibnamefont
  {F{\"a}hnle}},\ }\bibfield  {title} {\bibinfo {title} {Spin-cluster
  expansion: {{Parametrization}} of the general adiabatic magnetic energy
  surface with {\emph{ab initio}} accuracy},\ }\href
  {https://doi.org/10.1103/PhysRevB.69.104404} {\bibfield  {journal} {\bibinfo
  {journal} {Physical Review B}\ }\textbf {\bibinfo {volume} {69}},\ \bibinfo
  {pages} {104404} (\bibinfo {year} {2004})}\BibitemShut {NoStop}%
\bibitem [{\citenamefont {Van De~Walle}(2008)}]{vandewalle2008}%
  \BibitemOpen
  \bibfield  {author} {\bibinfo {author} {\bibfnamefont {A.}~\bibnamefont {Van
  De~Walle}},\ }\bibfield  {title} {\bibinfo {title} {A complete representation
  of structure--property relationships in crystals},\ }\href
  {https://doi.org/10.1038/nmat2200} {\bibfield  {journal} {\bibinfo  {journal}
  {Nature Materials}\ }\textbf {\bibinfo {volume} {7}},\ \bibinfo {pages} {455}
  (\bibinfo {year} {2008})}\BibitemShut {NoStop}%
\bibitem [{\citenamefont {Natarajan}\ and\ \citenamefont {{Van der
  Ven}}(2020)}]{natarajan2020a}%
  \BibitemOpen
  \bibfield  {author} {\bibinfo {author} {\bibfnamefont {A.~R.}\ \bibnamefont
  {Natarajan}}\ and\ \bibinfo {author} {\bibfnamefont {A.}~\bibnamefont {{Van
  der Ven}}},\ }\bibfield  {title} {\bibinfo {title} {Linking electronic
  structure calculations to generalized stacking fault energies in
  multicomponent alloys},\ }\href {https://doi.org/10.1038/s41524-020-0348-z}
  {\bibfield  {journal} {\bibinfo  {journal} {npj Computational Materials}\
  }\textbf {\bibinfo {volume} {6}},\ \bibinfo {pages} {80} (\bibinfo {year}
  {2020})}\BibitemShut {NoStop}%
\bibitem [{\citenamefont {{Van der Ven}}\ \emph {et~al.}(2001)\citenamefont
  {{Van der Ven}}, \citenamefont {Ceder}, \citenamefont {Asta},\ and\
  \citenamefont {Tepesch}}]{vanderven2001}%
  \BibitemOpen
  \bibfield  {author} {\bibinfo {author} {\bibfnamefont {A.}~\bibnamefont {{Van
  der Ven}}}, \bibinfo {author} {\bibfnamefont {G.}~\bibnamefont {Ceder}},
  \bibinfo {author} {\bibfnamefont {M.}~\bibnamefont {Asta}},\ and\ \bibinfo
  {author} {\bibfnamefont {P.~D.}\ \bibnamefont {Tepesch}},\ }\bibfield
  {title} {\bibinfo {title} {First-principles theory of ionic diffusion with
  nondilute carriers},\ }\href {https://doi.org/10.1103/PhysRevB.64.184307}
  {\bibfield  {journal} {\bibinfo  {journal} {Physical Review B}\ }\textbf
  {\bibinfo {volume} {64}},\ \bibinfo {pages} {184307} (\bibinfo {year}
  {2001})}\BibitemShut {NoStop}%
\bibitem [{\citenamefont {Behara}\ \emph {et~al.}(2024)\citenamefont {Behara},
  \citenamefont {Thomas}, \citenamefont {Puchala},\ and\ \citenamefont {Van
  Der~Ven}}]{behara2024a}%
  \BibitemOpen
  \bibfield  {author} {\bibinfo {author} {\bibfnamefont {S.~S.}\ \bibnamefont
  {Behara}}, \bibinfo {author} {\bibfnamefont {J.~C.}\ \bibnamefont {Thomas}},
  \bibinfo {author} {\bibfnamefont {B.}~\bibnamefont {Puchala}},\ and\ \bibinfo
  {author} {\bibfnamefont {A.}~\bibnamefont {Van Der~Ven}},\ }\bibfield
  {title} {\bibinfo {title} {Chemomechanics in alloy phase stability},\ }\href
  {https://doi.org/10.1103/PhysRevMaterials.8.033801} {\bibfield  {journal}
  {\bibinfo  {journal} {Physical Review Materials}\ }\textbf {\bibinfo {volume}
  {8}},\ \bibinfo {pages} {033801} (\bibinfo {year} {2024})}\BibitemShut
  {NoStop}%
\bibitem [{\citenamefont {Blum}\ \emph {et~al.}(2005)\citenamefont {Blum},
  \citenamefont {Hart}, \citenamefont {Walorski},\ and\ \citenamefont
  {Zunger}}]{blum2005}%
  \BibitemOpen
  \bibfield  {author} {\bibinfo {author} {\bibfnamefont {V.}~\bibnamefont
  {Blum}}, \bibinfo {author} {\bibfnamefont {G.~L.~W.}\ \bibnamefont {Hart}},
  \bibinfo {author} {\bibfnamefont {M.~J.}\ \bibnamefont {Walorski}},\ and\
  \bibinfo {author} {\bibfnamefont {A.}~\bibnamefont {Zunger}},\ }\bibfield
  {title} {\bibinfo {title} {Using genetic algorithms to map first-principles
  results to model {{Hamiltonians}}: {{Application}} to the generalized
  {{Ising}} model for alloys},\ }\href
  {https://doi.org/10.1103/PhysRevB.72.165113} {\bibfield  {journal} {\bibinfo
  {journal} {Physical Review B}\ }\textbf {\bibinfo {volume} {72}},\ \bibinfo
  {pages} {165113} (\bibinfo {year} {2005})}\BibitemShut {NoStop}%
\bibitem [{\citenamefont {Puchala}\ and\ \citenamefont {Van~der
  Ven}(2013)}]{puchala2013}%
  \BibitemOpen
  \bibfield  {author} {\bibinfo {author} {\bibfnamefont {B.}~\bibnamefont
  {Puchala}}\ and\ \bibinfo {author} {\bibfnamefont {A.}~\bibnamefont {Van~der
  Ven}},\ }\bibfield  {title} {\bibinfo {title} {Thermodynamics of the zr-o
  system from first-principles calculations},\ }\href
  {https://doi.org/10.1103/PhysRevB.88.094108} {\bibfield  {journal} {\bibinfo
  {journal} {Phys. Rev. B}\ }\textbf {\bibinfo {volume} {88}},\ \bibinfo
  {pages} {094108} (\bibinfo {year} {2013})}\BibitemShut {NoStop}%
\bibitem [{\citenamefont {Walle}\ and\ \citenamefont
  {Ceder}(2002)}]{walle2002}%
  \BibitemOpen
  \bibfield  {author} {\bibinfo {author} {\bibfnamefont {A.}~\bibnamefont
  {Walle}}\ and\ \bibinfo {author} {\bibfnamefont {G.}~\bibnamefont {Ceder}},\
  }\bibfield  {title} {\bibinfo {title} {Automating first-principles phase
  diagram calculations},\ }\href {https://doi.org/10.1361/105497102770331596}
  {\bibfield  {journal} {\bibinfo  {journal} {Journal of Phase Equilibria}\
  }\textbf {\bibinfo {volume} {23}},\ \bibinfo {pages} {348} (\bibinfo {year}
  {2002})}\BibitemShut {NoStop}%
\bibitem [{\citenamefont {Nelson}\ \emph {et~al.}(2013)\citenamefont {Nelson},
  \citenamefont {Hart}, \citenamefont {Zhou},\ and\ \citenamefont {Ozoli{\c
  n}{\v s}}}]{nelson2013}%
  \BibitemOpen
  \bibfield  {author} {\bibinfo {author} {\bibfnamefont {L.~J.}\ \bibnamefont
  {Nelson}}, \bibinfo {author} {\bibfnamefont {G.~L.~W.}\ \bibnamefont {Hart}},
  \bibinfo {author} {\bibfnamefont {F.}~\bibnamefont {Zhou}},\ and\ \bibinfo
  {author} {\bibfnamefont {V.}~\bibnamefont {Ozoli{\c n}{\v s}}},\ }\bibfield
  {title} {\bibinfo {title} {Compressive sensing as a paradigm for building
  physics models},\ }\href {https://doi.org/10.1103/PhysRevB.87.035125}
  {\bibfield  {journal} {\bibinfo  {journal} {Physical Review B}\ }\textbf
  {\bibinfo {volume} {87}},\ \bibinfo {pages} {035125} (\bibinfo {year}
  {2013})}\BibitemShut {NoStop}%
\bibitem [{\citenamefont {Mueller}\ and\ \citenamefont
  {Ceder}(2009)}]{mueller2009}%
  \BibitemOpen
  \bibfield  {author} {\bibinfo {author} {\bibfnamefont {T.}~\bibnamefont
  {Mueller}}\ and\ \bibinfo {author} {\bibfnamefont {G.}~\bibnamefont
  {Ceder}},\ }\bibfield  {title} {\bibinfo {title} {Bayesian approach to
  cluster expansions},\ }\href {https://doi.org/10.1103/PhysRevB.80.024103}
  {\bibfield  {journal} {\bibinfo  {journal} {Physical Review B}\ }\textbf
  {\bibinfo {volume} {80}},\ \bibinfo {pages} {024103} (\bibinfo {year}
  {2009})}\BibitemShut {NoStop}%
\bibitem [{\citenamefont {{Barroso-Luque}}\ \emph {et~al.}(2021)\citenamefont
  {{Barroso-Luque}}, \citenamefont {Yang},\ and\ \citenamefont
  {Ceder}}]{barroso-luque2021}%
  \BibitemOpen
  \bibfield  {author} {\bibinfo {author} {\bibfnamefont {L.}~\bibnamefont
  {{Barroso-Luque}}}, \bibinfo {author} {\bibfnamefont {J.~H.}\ \bibnamefont
  {Yang}},\ and\ \bibinfo {author} {\bibfnamefont {G.}~\bibnamefont {Ceder}},\
  }\bibfield  {title} {\bibinfo {title} {Sparse expansions of multicomponent
  oxide configuration energy using coherency and redundancy},\ }\href
  {https://doi.org/10.1103/PhysRevB.104.224203} {\bibfield  {journal} {\bibinfo
   {journal} {Physical Review B}\ }\textbf {\bibinfo {volume} {104}},\ \bibinfo
  {pages} {224203} (\bibinfo {year} {2021})}\BibitemShut {NoStop}%
\bibitem [{\citenamefont {{Barroso-Luque}}\ \emph {et~al.}(2022)\citenamefont
  {{Barroso-Luque}}, \citenamefont {Zhong}, \citenamefont {Yang}, \citenamefont
  {Xie}, \citenamefont {Chen}, \citenamefont {Ouyang},\ and\ \citenamefont
  {Ceder}}]{barroso-luque2022}%
  \BibitemOpen
  \bibfield  {author} {\bibinfo {author} {\bibfnamefont {L.}~\bibnamefont
  {{Barroso-Luque}}}, \bibinfo {author} {\bibfnamefont {P.}~\bibnamefont
  {Zhong}}, \bibinfo {author} {\bibfnamefont {J.~H.}\ \bibnamefont {Yang}},
  \bibinfo {author} {\bibfnamefont {F.}~\bibnamefont {Xie}}, \bibinfo {author}
  {\bibfnamefont {T.}~\bibnamefont {Chen}}, \bibinfo {author} {\bibfnamefont
  {B.}~\bibnamefont {Ouyang}},\ and\ \bibinfo {author} {\bibfnamefont
  {G.}~\bibnamefont {Ceder}},\ }\bibfield  {title} {\bibinfo {title} {Cluster
  expansions of multicomponent ionic materials: {{Formalism}} and
  methodology},\ }\href {https://doi.org/10.1103/PhysRevB.106.144202}
  {\bibfield  {journal} {\bibinfo  {journal} {Physical Review B}\ }\textbf
  {\bibinfo {volume} {106}},\ \bibinfo {pages} {144202} (\bibinfo {year}
  {2022})}\BibitemShut {NoStop}%
\bibitem [{\citenamefont {Zhong}\ \emph {et~al.}(2022)\citenamefont {Zhong},
  \citenamefont {Chen}, \citenamefont {{Barroso-Luque}}, \citenamefont {Xie},\
  and\ \citenamefont {Ceder}}]{zhong2022}%
  \BibitemOpen
  \bibfield  {author} {\bibinfo {author} {\bibfnamefont {P.}~\bibnamefont
  {Zhong}}, \bibinfo {author} {\bibfnamefont {T.}~\bibnamefont {Chen}},
  \bibinfo {author} {\bibfnamefont {L.}~\bibnamefont {{Barroso-Luque}}},
  \bibinfo {author} {\bibfnamefont {F.}~\bibnamefont {Xie}},\ and\ \bibinfo
  {author} {\bibfnamefont {G.}~\bibnamefont {Ceder}},\ }\bibfield  {title}
  {\bibinfo {title} {An {$\ell$} 0 {$\ell$} 2 -norm regularized regression
  model for construction of robust cluster expansions in multicomponent
  systems},\ }\href {https://doi.org/10.1103/PhysRevB.106.024203} {\bibfield
  {journal} {\bibinfo  {journal} {Physical Review B}\ }\textbf {\bibinfo
  {volume} {106}},\ \bibinfo {pages} {024203} (\bibinfo {year}
  {2022})}\BibitemShut {NoStop}%
\bibitem [{\citenamefont {{Barroso-Luque}}\ and\ \citenamefont
  {Ceder}(2024)}]{barroso-luque2024}%
  \BibitemOpen
  \bibfield  {author} {\bibinfo {author} {\bibfnamefont {L.}~\bibnamefont
  {{Barroso-Luque}}}\ and\ \bibinfo {author} {\bibfnamefont {G.}~\bibnamefont
  {Ceder}},\ }\bibfield  {title} {\bibinfo {title} {The cluster decomposition
  of the configurational energy of multicomponent alloys},\ }\href
  {https://doi.org/10.1038/s41524-024-01338-y} {\bibfield  {journal} {\bibinfo
  {journal} {npj Computational Materials}\ }\textbf {\bibinfo {volume} {10}},\
  \bibinfo {pages} {158} (\bibinfo {year} {2024})}\BibitemShut {NoStop}%
\bibitem [{\citenamefont {Kadkhodaei}\ and\ \citenamefont
  {Mu{\~n}oz}(2021)}]{kadkhodaei2021}%
  \BibitemOpen
  \bibfield  {author} {\bibinfo {author} {\bibfnamefont {S.}~\bibnamefont
  {Kadkhodaei}}\ and\ \bibinfo {author} {\bibfnamefont {J.~A.}\ \bibnamefont
  {Mu{\~n}oz}},\ }\bibfield  {title} {\bibinfo {title} {Cluster {{Expansion}}
  of {{Alloy Theory}}: {{A Review}} of {{Historical Development}} and {{Modern
  Innovations}}},\ }\href {https://doi.org/10.1007/s11837-021-04840-6}
  {\bibfield  {journal} {\bibinfo  {journal} {JOM}\ }\textbf {\bibinfo {volume}
  {73}},\ \bibinfo {pages} {3326} (\bibinfo {year} {2021})}\BibitemShut
  {NoStop}%
\bibitem [{\citenamefont {Aldegunde}\ \emph {et~al.}(2016)\citenamefont
  {Aldegunde}, \citenamefont {Zabaras},\ and\ \citenamefont
  {Kristensen}}]{aldegunde2016}%
  \BibitemOpen
  \bibfield  {author} {\bibinfo {author} {\bibfnamefont {M.}~\bibnamefont
  {Aldegunde}}, \bibinfo {author} {\bibfnamefont {N.}~\bibnamefont {Zabaras}},\
  and\ \bibinfo {author} {\bibfnamefont {J.}~\bibnamefont {Kristensen}},\
  }\bibfield  {title} {\bibinfo {title} {Quantifying uncertainties in
  first-principles alloy thermodynamics using cluster expansions},\ }\href
  {https://doi.org/10.1016/j.jcp.2016.07.016} {\bibfield  {journal} {\bibinfo
  {journal} {Journal of Computational Physics}\ }\textbf {\bibinfo {volume}
  {323}},\ \bibinfo {pages} {17} (\bibinfo {year} {2016})}\BibitemShut
  {NoStop}%
\bibitem [{\citenamefont {Kristensen}\ and\ \citenamefont
  {Zabaras}(2014)}]{kristensen2014}%
  \BibitemOpen
  \bibfield  {author} {\bibinfo {author} {\bibfnamefont {J.}~\bibnamefont
  {Kristensen}}\ and\ \bibinfo {author} {\bibfnamefont {N.~J.}\ \bibnamefont
  {Zabaras}},\ }\bibfield  {title} {\bibinfo {title} {Bayesian uncertainty
  quantification in the evaluation of alloy properties with the cluster
  expansion method},\ }\href {https://doi.org/10.1016/j.cpc.2014.07.013}
  {\bibfield  {journal} {\bibinfo  {journal} {Computer Physics Communications}\
  }\textbf {\bibinfo {volume} {185}},\ \bibinfo {pages} {2885} (\bibinfo {year}
  {2014})}\BibitemShut {NoStop}%
\bibitem [{\citenamefont {Ober}\ and\ \citenamefont {{Van der
  Ven}}(2023)}]{ober2023}%
  \BibitemOpen
  \bibfield  {author} {\bibinfo {author} {\bibfnamefont {D.~E.}\ \bibnamefont
  {Ober}}\ and\ \bibinfo {author} {\bibfnamefont {A.}~\bibnamefont {{Van der
  Ven}}},\ }\href@noop {} {\bibinfo {title} {Thermodynamically {{Informed
  Priors}} for {{Uncertainty Propagation}} in {{First-Principles Statistical
  Mechanics}}}} (\bibinfo {year} {2023}),\ \Eprint
  {https://arxiv.org/abs/2309.12255} {arXiv:2309.12255 [cond-mat]} \BibitemShut
  {NoStop}%
\bibitem [{\citenamefont {Wen}\ \emph {et~al.}(2023)\citenamefont {Wen},
  \citenamefont {Tucker},\ and\ \citenamefont {Titus}}]{wen2023}%
  \BibitemOpen
  \bibfield  {author} {\bibinfo {author} {\bibfnamefont {D.}~\bibnamefont
  {Wen}}, \bibinfo {author} {\bibfnamefont {V.}~\bibnamefont {Tucker}},\ and\
  \bibinfo {author} {\bibfnamefont {M.}~\bibnamefont {Titus}},\ }\href
  {https://doi.org/10.21203/rs.3.rs-3649931/v1} {\bibinfo {title} {Bayesian
  {{Optimization Acquisition Functions}} for {{Accelerated Search}} of {{Energy
  Convex Hull}} of {{Multi-Component Alloys}}}} (\bibinfo {year}
  {2023})\BibitemShut {NoStop}%
\bibitem [{\citenamefont {Chen}\ \emph {et~al.}(2024)\citenamefont {Chen},
  \citenamefont {Samanta}, \citenamefont {Zhu}, \citenamefont {Eckert},
  \citenamefont {Schroers}, \citenamefont {Curtarolo},\ and\ \citenamefont {Van
  De~Walle}}]{chen2024d}%
  \BibitemOpen
  \bibfield  {author} {\bibinfo {author} {\bibfnamefont {H.}~\bibnamefont
  {Chen}}, \bibinfo {author} {\bibfnamefont {S.}~\bibnamefont {Samanta}},
  \bibinfo {author} {\bibfnamefont {S.}~\bibnamefont {Zhu}}, \bibinfo {author}
  {\bibfnamefont {H.}~\bibnamefont {Eckert}}, \bibinfo {author} {\bibfnamefont
  {J.}~\bibnamefont {Schroers}}, \bibinfo {author} {\bibfnamefont
  {S.}~\bibnamefont {Curtarolo}},\ and\ \bibinfo {author} {\bibfnamefont
  {A.}~\bibnamefont {Van De~Walle}},\ }\bibfield  {title} {\bibinfo {title}
  {Bayesian active machine learning for {{Cluster}} expansion construction},\
  }\href {https://doi.org/10.1016/j.commatsci.2023.112571} {\bibfield
  {journal} {\bibinfo  {journal} {Computational Materials Science}\ }\textbf
  {\bibinfo {volume} {231}},\ \bibinfo {pages} {112571} (\bibinfo {year}
  {2024})}\BibitemShut {NoStop}%
\bibitem [{\citenamefont {Natarajan}\ \emph {et~al.}(2016)\citenamefont
  {Natarajan}, \citenamefont {Solomon}, \citenamefont {Puchala}, \citenamefont
  {Marquis},\ and\ \citenamefont {{Van der Ven}}}]{natarajan2016}%
  \BibitemOpen
  \bibfield  {author} {\bibinfo {author} {\bibfnamefont {A.~R.}\ \bibnamefont
  {Natarajan}}, \bibinfo {author} {\bibfnamefont {E.~L.}\ \bibnamefont
  {Solomon}}, \bibinfo {author} {\bibfnamefont {B.}~\bibnamefont {Puchala}},
  \bibinfo {author} {\bibfnamefont {E.~A.}\ \bibnamefont {Marquis}},\ and\
  \bibinfo {author} {\bibfnamefont {A.}~\bibnamefont {{Van der Ven}}},\
  }\bibfield  {title} {\bibinfo {title} {On the early stages of precipitation
  in dilute {{Mg}}--{{Nd}} alloys},\ }\href
  {https://doi.org/10.1016/j.actamat.2016.01.055} {\bibfield  {journal}
  {\bibinfo  {journal} {Acta Materialia}\ }\textbf {\bibinfo {volume} {108}},\
  \bibinfo {pages} {367} (\bibinfo {year} {2016})}\BibitemShut {NoStop}%
\bibitem [{\citenamefont {Natarajan}\ and\ \citenamefont {{Van der
  Ven}}(2017{\natexlab{b}})}]{natarajan2017}%
  \BibitemOpen
  \bibfield  {author} {\bibinfo {author} {\bibfnamefont {A.~R.}\ \bibnamefont
  {Natarajan}}\ and\ \bibinfo {author} {\bibfnamefont {A.}~\bibnamefont {{Van
  der Ven}}},\ }\bibfield  {title} {\bibinfo {title} {A unified description of
  ordering in {{HCP Mg-RE}} alloys},\ }\href
  {https://doi.org/10.1016/j.actamat.2016.10.057} {\bibfield  {journal}
  {\bibinfo  {journal} {Acta Materialia}\ }\textbf {\bibinfo {volume} {124}},\
  \bibinfo {pages} {620} (\bibinfo {year} {2017}{\natexlab{b}})}\BibitemShut
  {NoStop}%
\bibitem [{\citenamefont {Smith}\ \emph {et~al.}(2024)\citenamefont {Smith},
  \citenamefont {Liu}, \citenamefont {Xia},\ and\ \citenamefont
  {Wolverton}}]{smith2024}%
  \BibitemOpen
  \bibfield  {author} {\bibinfo {author} {\bibfnamefont {N.~C.}\ \bibnamefont
  {Smith}}, \bibinfo {author} {\bibfnamefont {T.-c.}\ \bibnamefont {Liu}},
  \bibinfo {author} {\bibfnamefont {Y.}~\bibnamefont {Xia}},\ and\ \bibinfo
  {author} {\bibfnamefont {C.}~\bibnamefont {Wolverton}},\ }\bibfield  {title}
  {\bibinfo {title} {Competition between long- and short-range order in
  size-mismatched medium-entropy alloys},\ }\href
  {https://doi.org/10.1016/j.actamat.2024.120199} {\bibfield  {journal}
  {\bibinfo  {journal} {Acta Materialia}\ }\textbf {\bibinfo {volume} {277}},\
  \bibinfo {pages} {120199} (\bibinfo {year} {2024})}\BibitemShut {NoStop}%
\bibitem [{\citenamefont {Huang}\ \emph {et~al.}(2015)\citenamefont {Huang},
  \citenamefont {Zhao}, \citenamefont {Cao}, \citenamefont {Chen},
  \citenamefont {Zhu}, \citenamefont {Lin}, \citenamefont {Li}, \citenamefont
  {Yan}, \citenamefont {Zettl}, \citenamefont {Wang}, \citenamefont {Duan},
  \citenamefont {Mueller},\ and\ \citenamefont {Huang}}]{huang2015}%
  \BibitemOpen
  \bibfield  {author} {\bibinfo {author} {\bibfnamefont {X.}~\bibnamefont
  {Huang}}, \bibinfo {author} {\bibfnamefont {Z.}~\bibnamefont {Zhao}},
  \bibinfo {author} {\bibfnamefont {L.}~\bibnamefont {Cao}}, \bibinfo {author}
  {\bibfnamefont {Y.}~\bibnamefont {Chen}}, \bibinfo {author} {\bibfnamefont
  {E.}~\bibnamefont {Zhu}}, \bibinfo {author} {\bibfnamefont {Z.}~\bibnamefont
  {Lin}}, \bibinfo {author} {\bibfnamefont {M.}~\bibnamefont {Li}}, \bibinfo
  {author} {\bibfnamefont {A.}~\bibnamefont {Yan}}, \bibinfo {author}
  {\bibfnamefont {A.}~\bibnamefont {Zettl}}, \bibinfo {author} {\bibfnamefont
  {Y.~M.}\ \bibnamefont {Wang}}, \bibinfo {author} {\bibfnamefont
  {X.}~\bibnamefont {Duan}}, \bibinfo {author} {\bibfnamefont {T.}~\bibnamefont
  {Mueller}},\ and\ \bibinfo {author} {\bibfnamefont {Y.}~\bibnamefont
  {Huang}},\ }\bibfield  {title} {\bibinfo {title} {High-performance transition
  metal--doped {{Pt}} {\textsubscript{3}} {{Ni}} octahedra for oxygen reduction
  reaction},\ }\href {https://doi.org/10.1126/science.aaa8765} {\bibfield
  {journal} {\bibinfo  {journal} {Science}\ }\textbf {\bibinfo {volume}
  {348}},\ \bibinfo {pages} {1230} (\bibinfo {year} {2015})}\BibitemShut
  {NoStop}%
\bibitem [{\citenamefont {Cao}\ \emph {et~al.}(2019)\citenamefont {Cao},
  \citenamefont {Niu},\ and\ \citenamefont {Mueller}}]{cao2019}%
  \BibitemOpen
  \bibfield  {author} {\bibinfo {author} {\bibfnamefont {L.}~\bibnamefont
  {Cao}}, \bibinfo {author} {\bibfnamefont {L.}~\bibnamefont {Niu}},\ and\
  \bibinfo {author} {\bibfnamefont {T.}~\bibnamefont {Mueller}},\ }\bibfield
  {title} {\bibinfo {title} {Computationally generated maps of surface
  structures and catalytic activities for alloy phase diagrams},\ }\href
  {https://doi.org/10.1073/pnas.1910724116} {\bibfield  {journal} {\bibinfo
  {journal} {Proceedings of the National Academy of Sciences}\ }\textbf
  {\bibinfo {volume} {116}},\ \bibinfo {pages} {22044} (\bibinfo {year}
  {2019})}\BibitemShut {NoStop}%
\bibitem [{\citenamefont {Kitchaev}\ \emph {et~al.}(2018)\citenamefont
  {Kitchaev}, \citenamefont {Lun}, \citenamefont {Richards}, \citenamefont
  {Ji}, \citenamefont {Cl{\'e}ment}, \citenamefont {Balasubramanian},
  \citenamefont {Kwon}, \citenamefont {Dai}, \citenamefont {Papp},
  \citenamefont {Lei}, \citenamefont {McCloskey}, \citenamefont {Yang},
  \citenamefont {Lee},\ and\ \citenamefont {Ceder}}]{kitchaev2018}%
  \BibitemOpen
  \bibfield  {author} {\bibinfo {author} {\bibfnamefont {D.~A.}\ \bibnamefont
  {Kitchaev}}, \bibinfo {author} {\bibfnamefont {Z.}~\bibnamefont {Lun}},
  \bibinfo {author} {\bibfnamefont {W.~D.}\ \bibnamefont {Richards}}, \bibinfo
  {author} {\bibfnamefont {H.}~\bibnamefont {Ji}}, \bibinfo {author}
  {\bibfnamefont {R.~J.}\ \bibnamefont {Cl{\'e}ment}}, \bibinfo {author}
  {\bibfnamefont {M.}~\bibnamefont {Balasubramanian}}, \bibinfo {author}
  {\bibfnamefont {D.-H.}\ \bibnamefont {Kwon}}, \bibinfo {author}
  {\bibfnamefont {K.}~\bibnamefont {Dai}}, \bibinfo {author} {\bibfnamefont
  {J.~K.}\ \bibnamefont {Papp}}, \bibinfo {author} {\bibfnamefont
  {T.}~\bibnamefont {Lei}}, \bibinfo {author} {\bibfnamefont {B.~D.}\
  \bibnamefont {McCloskey}}, \bibinfo {author} {\bibfnamefont {W.}~\bibnamefont
  {Yang}}, \bibinfo {author} {\bibfnamefont {J.}~\bibnamefont {Lee}},\ and\
  \bibinfo {author} {\bibfnamefont {G.}~\bibnamefont {Ceder}},\ }\bibfield
  {title} {\bibinfo {title} {Design principles for high transition metal
  capacity in disordered rocksalt {{Li-ion}} cathodes},\ }\href
  {https://doi.org/10.1039/C8EE00816G} {\bibfield  {journal} {\bibinfo
  {journal} {Energy \& Environmental Science}\ }\textbf {\bibinfo {volume}
  {11}},\ \bibinfo {pages} {2159} (\bibinfo {year} {2018})}\BibitemShut
  {NoStop}%
\bibitem [{\citenamefont {Richards}\ \emph {et~al.}(2018)\citenamefont
  {Richards}, \citenamefont {Dacek}, \citenamefont {Kitchaev},\ and\
  \citenamefont {Ceder}}]{richards2018}%
  \BibitemOpen
  \bibfield  {author} {\bibinfo {author} {\bibfnamefont {W.~D.}\ \bibnamefont
  {Richards}}, \bibinfo {author} {\bibfnamefont {S.~T.}\ \bibnamefont {Dacek}},
  \bibinfo {author} {\bibfnamefont {D.~A.}\ \bibnamefont {Kitchaev}},\ and\
  \bibinfo {author} {\bibfnamefont {G.}~\bibnamefont {Ceder}},\ }\bibfield
  {title} {\bibinfo {title} {Fluorination of {{Lithium}}-{{Excess Transition
  Metal Oxide Cathode Materials}}},\ }\href
  {https://doi.org/10.1002/aenm.201701533} {\bibfield  {journal} {\bibinfo
  {journal} {Advanced Energy Materials}\ }\textbf {\bibinfo {volume} {8}},\
  \bibinfo {pages} {1701533} (\bibinfo {year} {2018})}\BibitemShut {NoStop}%
\bibitem [{\citenamefont {Lun}\ \emph {et~al.}(2019)\citenamefont {Lun},
  \citenamefont {Ouyang}, \citenamefont {Kitchaev}, \citenamefont
  {Cl{\'e}ment}, \citenamefont {Papp}, \citenamefont {Balasubramanian},
  \citenamefont {Tian}, \citenamefont {Lei}, \citenamefont {Shi}, \citenamefont
  {McCloskey}, \citenamefont {Lee},\ and\ \citenamefont {Ceder}}]{lun2019}%
  \BibitemOpen
  \bibfield  {author} {\bibinfo {author} {\bibfnamefont {Z.}~\bibnamefont
  {Lun}}, \bibinfo {author} {\bibfnamefont {B.}~\bibnamefont {Ouyang}},
  \bibinfo {author} {\bibfnamefont {D.~A.}\ \bibnamefont {Kitchaev}}, \bibinfo
  {author} {\bibfnamefont {R.~J.}\ \bibnamefont {Cl{\'e}ment}}, \bibinfo
  {author} {\bibfnamefont {J.~K.}\ \bibnamefont {Papp}}, \bibinfo {author}
  {\bibfnamefont {M.}~\bibnamefont {Balasubramanian}}, \bibinfo {author}
  {\bibfnamefont {Y.}~\bibnamefont {Tian}}, \bibinfo {author} {\bibfnamefont
  {T.}~\bibnamefont {Lei}}, \bibinfo {author} {\bibfnamefont {T.}~\bibnamefont
  {Shi}}, \bibinfo {author} {\bibfnamefont {B.~D.}\ \bibnamefont {McCloskey}},
  \bibinfo {author} {\bibfnamefont {J.}~\bibnamefont {Lee}},\ and\ \bibinfo
  {author} {\bibfnamefont {G.}~\bibnamefont {Ceder}},\ }\bibfield  {title}
  {\bibinfo {title} {Improved {{Cycling Performance}} of {{Li}}-{{Excess
  Cation}}-{{Disordered Cathode Materials}} upon {{Fluorine Substitution}}},\
  }\href {https://doi.org/10.1002/aenm.201802959} {\bibfield  {journal}
  {\bibinfo  {journal} {Advanced Energy Materials}\ }\textbf {\bibinfo {volume}
  {9}},\ \bibinfo {pages} {1802959} (\bibinfo {year} {2019})}\BibitemShut
  {NoStop}%
\bibitem [{\citenamefont {Lun}\ \emph {et~al.}(2021)\citenamefont {Lun},
  \citenamefont {Ouyang}, \citenamefont {Kwon}, \citenamefont {Ha},
  \citenamefont {Foley}, \citenamefont {Huang}, \citenamefont {Cai},
  \citenamefont {Kim}, \citenamefont {Balasubramanian}, \citenamefont {Sun},
  \citenamefont {Huang}, \citenamefont {Tian}, \citenamefont {Kim},
  \citenamefont {McCloskey}, \citenamefont {Yang}, \citenamefont {Cl{\'e}ment},
  \citenamefont {Ji},\ and\ \citenamefont {Ceder}}]{lun2021}%
  \BibitemOpen
  \bibfield  {author} {\bibinfo {author} {\bibfnamefont {Z.}~\bibnamefont
  {Lun}}, \bibinfo {author} {\bibfnamefont {B.}~\bibnamefont {Ouyang}},
  \bibinfo {author} {\bibfnamefont {D.-H.}\ \bibnamefont {Kwon}}, \bibinfo
  {author} {\bibfnamefont {Y.}~\bibnamefont {Ha}}, \bibinfo {author}
  {\bibfnamefont {E.~E.}\ \bibnamefont {Foley}}, \bibinfo {author}
  {\bibfnamefont {T.-Y.}\ \bibnamefont {Huang}}, \bibinfo {author}
  {\bibfnamefont {Z.}~\bibnamefont {Cai}}, \bibinfo {author} {\bibfnamefont
  {H.}~\bibnamefont {Kim}}, \bibinfo {author} {\bibfnamefont {M.}~\bibnamefont
  {Balasubramanian}}, \bibinfo {author} {\bibfnamefont {Y.}~\bibnamefont
  {Sun}}, \bibinfo {author} {\bibfnamefont {J.}~\bibnamefont {Huang}}, \bibinfo
  {author} {\bibfnamefont {Y.}~\bibnamefont {Tian}}, \bibinfo {author}
  {\bibfnamefont {H.}~\bibnamefont {Kim}}, \bibinfo {author} {\bibfnamefont
  {B.~D.}\ \bibnamefont {McCloskey}}, \bibinfo {author} {\bibfnamefont
  {W.}~\bibnamefont {Yang}}, \bibinfo {author} {\bibfnamefont {R.~J.}\
  \bibnamefont {Cl{\'e}ment}}, \bibinfo {author} {\bibfnamefont
  {H.}~\bibnamefont {Ji}},\ and\ \bibinfo {author} {\bibfnamefont
  {G.}~\bibnamefont {Ceder}},\ }\bibfield  {title} {\bibinfo {title}
  {Cation-disordered rocksalt-type high-entropy cathodes for {{Li-ion}}
  batteries},\ }\href {https://doi.org/10.1038/s41563-020-00816-0} {\bibfield
  {journal} {\bibinfo  {journal} {Nature Materials}\ }\textbf {\bibinfo
  {volume} {20}},\ \bibinfo {pages} {214} (\bibinfo {year} {2021})}\BibitemShut
  {NoStop}%
\bibitem [{\citenamefont {Zunger}(1994)}]{zunger1994}%
  \BibitemOpen
  \bibfield  {author} {\bibinfo {author} {\bibfnamefont {A.}~\bibnamefont
  {Zunger}},\ }\bibfield  {title} {\bibinfo {title} {First-{{Principles
  Statistical Mechanics}} of {{Semiconductor Alloys}} and {{Intermetallic
  Compounds}}},\ }in\ \href {https://doi.org/10.1007/978-1-4615-2476-2_23}
  {\emph {\bibinfo {booktitle} {Statics and {{Dynamics}} of {{Alloy Phase
  Transformations}}}}},\ Vol.\ \bibinfo {volume} {319},\ \bibinfo {editor}
  {edited by\ \bibinfo {editor} {\bibfnamefont {P.~E.~A.}\ \bibnamefont
  {Turchi}}\ and\ \bibinfo {editor} {\bibfnamefont {A.}~\bibnamefont {Gonis}}}\
  (\bibinfo  {publisher} {Springer US},\ \bibinfo {address} {Boston, MA},\
  \bibinfo {year} {1994})\ pp.\ \bibinfo {pages} {361--419}\BibitemShut
  {NoStop}%
\bibitem [{\citenamefont {George}\ \emph {et~al.}(2019)\citenamefont {George},
  \citenamefont {Raabe},\ and\ \citenamefont {Ritchie}}]{george2019}%
  \BibitemOpen
  \bibfield  {author} {\bibinfo {author} {\bibfnamefont {E.~P.}\ \bibnamefont
  {George}}, \bibinfo {author} {\bibfnamefont {D.}~\bibnamefont {Raabe}},\ and\
  \bibinfo {author} {\bibfnamefont {R.~O.}\ \bibnamefont {Ritchie}},\
  }\bibfield  {title} {\bibinfo {title} {High-entropy alloys},\ }\href
  {https://doi.org/10.1038/s41578-019-0121-4} {\bibfield  {journal} {\bibinfo
  {journal} {Nature Reviews Materials}\ }\textbf {\bibinfo {volume} {4}},\
  \bibinfo {pages} {515} (\bibinfo {year} {2019})}\BibitemShut {NoStop}%
\bibitem [{\citenamefont {George}\ \emph {et~al.}(2020)\citenamefont {George},
  \citenamefont {Curtin},\ and\ \citenamefont {Tasan}}]{george2020}%
  \BibitemOpen
  \bibfield  {author} {\bibinfo {author} {\bibfnamefont {E.}~\bibnamefont
  {George}}, \bibinfo {author} {\bibfnamefont {W.}~\bibnamefont {Curtin}},\
  and\ \bibinfo {author} {\bibfnamefont {C.}~\bibnamefont {Tasan}},\ }\bibfield
   {title} {\bibinfo {title} {High entropy alloys: {{A}} focused review of
  mechanical properties and deformation mechanisms},\ }\href
  {https://doi.org/10.1016/j.actamat.2019.12.015} {\bibfield  {journal}
  {\bibinfo  {journal} {Acta Materialia}\ }\textbf {\bibinfo {volume} {188}},\
  \bibinfo {pages} {435} (\bibinfo {year} {2020})}\BibitemShut {NoStop}%
\bibitem [{\citenamefont {Miracle}\ \emph {et~al.}(2020)\citenamefont
  {Miracle}, \citenamefont {Tsai}, \citenamefont {Senkov}, \citenamefont
  {Soni},\ and\ \citenamefont {Banerjee}}]{miracle2020}%
  \BibitemOpen
  \bibfield  {author} {\bibinfo {author} {\bibfnamefont {D.~B.}\ \bibnamefont
  {Miracle}}, \bibinfo {author} {\bibfnamefont {M.-H.}\ \bibnamefont {Tsai}},
  \bibinfo {author} {\bibfnamefont {O.~N.}\ \bibnamefont {Senkov}}, \bibinfo
  {author} {\bibfnamefont {V.}~\bibnamefont {Soni}},\ and\ \bibinfo {author}
  {\bibfnamefont {R.}~\bibnamefont {Banerjee}},\ }\bibfield  {title} {\bibinfo
  {title} {Refractory high entropy superalloys ({{RSAs}})},\ }\href
  {https://doi.org/10.1016/j.scriptamat.2020.06.048} {\bibfield  {journal}
  {\bibinfo  {journal} {Scripta Materialia}\ }\textbf {\bibinfo {volume}
  {187}},\ \bibinfo {pages} {445} (\bibinfo {year} {2020})}\BibitemShut
  {NoStop}%
\bibitem [{\citenamefont {Miracle}\ \emph {et~al.}(2024)\citenamefont
  {Miracle}, \citenamefont {Senkov}, \citenamefont {Frey}, \citenamefont
  {Rao},\ and\ \citenamefont {Pollock}}]{miracle2024}%
  \BibitemOpen
  \bibfield  {author} {\bibinfo {author} {\bibfnamefont {D.}~\bibnamefont
  {Miracle}}, \bibinfo {author} {\bibfnamefont {O.}~\bibnamefont {Senkov}},
  \bibinfo {author} {\bibfnamefont {C.}~\bibnamefont {Frey}}, \bibinfo {author}
  {\bibfnamefont {S.}~\bibnamefont {Rao}},\ and\ \bibinfo {author}
  {\bibfnamefont {T.}~\bibnamefont {Pollock}},\ }\bibfield  {title} {\bibinfo
  {title} {Strength vs temperature for refractory complex concentrated alloys
  ({{RCCAs}}): {{A}} critical comparison with refractory {{BCC}} elements and
  dilute alloys},\ }\href {https://doi.org/10.1016/j.actamat.2024.119692}
  {\bibfield  {journal} {\bibinfo  {journal} {Acta Materialia}\ }\textbf
  {\bibinfo {volume} {266}},\ \bibinfo {pages} {119692} (\bibinfo {year}
  {2024})}\BibitemShut {NoStop}%
\bibitem [{\citenamefont {Schweidler}\ \emph {et~al.}(2024)\citenamefont
  {Schweidler}, \citenamefont {Botros}, \citenamefont {Strauss}, \citenamefont
  {Wang}, \citenamefont {Ma}, \citenamefont {Velasco}, \citenamefont
  {Cadilha~Marques}, \citenamefont {Sarkar}, \citenamefont {K{\"u}bel},
  \citenamefont {Hahn}, \citenamefont {{Aghassi-Hagmann}}, \citenamefont
  {Brezesinski},\ and\ \citenamefont {Breitung}}]{schweidler2024}%
  \BibitemOpen
  \bibfield  {author} {\bibinfo {author} {\bibfnamefont {S.}~\bibnamefont
  {Schweidler}}, \bibinfo {author} {\bibfnamefont {M.}~\bibnamefont {Botros}},
  \bibinfo {author} {\bibfnamefont {F.}~\bibnamefont {Strauss}}, \bibinfo
  {author} {\bibfnamefont {Q.}~\bibnamefont {Wang}}, \bibinfo {author}
  {\bibfnamefont {Y.}~\bibnamefont {Ma}}, \bibinfo {author} {\bibfnamefont
  {L.}~\bibnamefont {Velasco}}, \bibinfo {author} {\bibfnamefont
  {G.}~\bibnamefont {Cadilha~Marques}}, \bibinfo {author} {\bibfnamefont
  {A.}~\bibnamefont {Sarkar}}, \bibinfo {author} {\bibfnamefont
  {C.}~\bibnamefont {K{\"u}bel}}, \bibinfo {author} {\bibfnamefont
  {H.}~\bibnamefont {Hahn}}, \bibinfo {author} {\bibfnamefont {J.}~\bibnamefont
  {{Aghassi-Hagmann}}}, \bibinfo {author} {\bibfnamefont {T.}~\bibnamefont
  {Brezesinski}},\ and\ \bibinfo {author} {\bibfnamefont {B.}~\bibnamefont
  {Breitung}},\ }\bibfield  {title} {\bibinfo {title} {High-entropy materials
  for energy and electronic applications},\ }\href
  {https://doi.org/10.1038/s41578-024-00654-5} {\bibfield  {journal} {\bibinfo
  {journal} {Nature Reviews Materials}\ }\textbf {\bibinfo {volume} {9}},\
  \bibinfo {pages} {266} (\bibinfo {year} {2024})}\BibitemShut {NoStop}%
\bibitem [{\citenamefont {Sun}\ and\ \citenamefont {Dai}(2021)}]{sun2021a}%
  \BibitemOpen
  \bibfield  {author} {\bibinfo {author} {\bibfnamefont {Y.}~\bibnamefont
  {Sun}}\ and\ \bibinfo {author} {\bibfnamefont {S.}~\bibnamefont {Dai}},\
  }\bibfield  {title} {\bibinfo {title} {High-entropy materials for catalysis:
  {{A}} new frontier},\ }\href {https://doi.org/10.1126/sciadv.abg1600}
  {\bibfield  {journal} {\bibinfo  {journal} {Science Advances}\ }\textbf
  {\bibinfo {volume} {7}},\ \bibinfo {pages} {eabg1600} (\bibinfo {year}
  {2021})}\BibitemShut {NoStop}%
\bibitem [{\citenamefont {Van De~Walle}(2009)}]{vandewalle2009}%
  \BibitemOpen
  \bibfield  {author} {\bibinfo {author} {\bibfnamefont {A.}~\bibnamefont {Van
  De~Walle}},\ }\bibfield  {title} {\bibinfo {title} {Multicomponent
  multisublattice alloys, nonconfigurational entropy and other additions to the
  {{Alloy Theoretic Automated Toolkit}}},\ }\href
  {https://doi.org/10.1016/j.calphad.2008.12.005} {\bibfield  {journal}
  {\bibinfo  {journal} {Calphad}\ }\textbf {\bibinfo {volume} {33}},\ \bibinfo
  {pages} {266} (\bibinfo {year} {2009})}\BibitemShut {NoStop}%
\bibitem [{\citenamefont {Natarajan}\ and\ \citenamefont {{Van der
  Ven}}(2018)}]{natarajan2018}%
  \BibitemOpen
  \bibfield  {author} {\bibinfo {author} {\bibfnamefont {A.~R.}\ \bibnamefont
  {Natarajan}}\ and\ \bibinfo {author} {\bibfnamefont {A.}~\bibnamefont {{Van
  der Ven}}},\ }\bibfield  {title} {\bibinfo {title} {Machine-learning the
  configurational energy of multicomponent crystalline solids},\ }\href
  {https://doi.org/10.1038/s41524-018-0110-y} {\bibfield  {journal} {\bibinfo
  {journal} {npj Computational Materials}\ }\textbf {\bibinfo {volume} {4}},\
  \bibinfo {pages} {56} (\bibinfo {year} {2018})}\BibitemShut {NoStop}%
\bibitem [{\citenamefont {Natarajan}\ \emph {et~al.}(2020)\citenamefont
  {Natarajan}, \citenamefont {Dolin},\ and\ \citenamefont {{Van der
  Ven}}}]{natarajan2020}%
  \BibitemOpen
  \bibfield  {author} {\bibinfo {author} {\bibfnamefont {A.~R.}\ \bibnamefont
  {Natarajan}}, \bibinfo {author} {\bibfnamefont {P.}~\bibnamefont {Dolin}},\
  and\ \bibinfo {author} {\bibfnamefont {A.}~\bibnamefont {{Van der Ven}}},\
  }\bibfield  {title} {\bibinfo {title} {Crystallography, thermodynamics and
  phase transitions in refractory binary alloys},\ }\href
  {https://doi.org/10.1016/j.actamat.2020.08.034} {\bibfield  {journal}
  {\bibinfo  {journal} {Acta Materialia}\ }\textbf {\bibinfo {volume} {200}},\
  \bibinfo {pages} {171} (\bibinfo {year} {2020})}\BibitemShut {NoStop}%
\bibitem [{\citenamefont {Goiri}\ and\ \citenamefont {Van
  Der~Ven}(2018)}]{goiri2018}%
  \BibitemOpen
  \bibfield  {author} {\bibinfo {author} {\bibfnamefont {J.~G.}\ \bibnamefont
  {Goiri}}\ and\ \bibinfo {author} {\bibfnamefont {A.}~\bibnamefont {Van
  Der~Ven}},\ }\bibfield  {title} {\bibinfo {title} {Recursive alloy
  {{Hamiltonian}} construction and its application to the {{Ni-Al-Cr}}
  system},\ }\href {https://doi.org/10.1016/j.actamat.2018.06.048} {\bibfield
  {journal} {\bibinfo  {journal} {Acta Materialia}\ }\textbf {\bibinfo {volume}
  {159}},\ \bibinfo {pages} {257} (\bibinfo {year} {2018})}\BibitemShut
  {NoStop}%
\bibitem [{\citenamefont {Willatt}\ \emph {et~al.}(2018)\citenamefont
  {Willatt}, \citenamefont {Musil},\ and\ \citenamefont
  {Ceriotti}}]{willatt2018}%
  \BibitemOpen
  \bibfield  {author} {\bibinfo {author} {\bibfnamefont {M.~J.}\ \bibnamefont
  {Willatt}}, \bibinfo {author} {\bibfnamefont {F.}~\bibnamefont {Musil}},\
  and\ \bibinfo {author} {\bibfnamefont {M.}~\bibnamefont {Ceriotti}},\
  }\bibfield  {title} {\bibinfo {title} {Feature optimization for atomistic
  machine learning yields a data-driven construction of the periodic table of
  the elements},\ }\href {https://doi.org/10.1039/C8CP05921G} {\bibfield
  {journal} {\bibinfo  {journal} {Physical Chemistry Chemical Physics}\
  }\textbf {\bibinfo {volume} {20}},\ \bibinfo {pages} {29661} (\bibinfo {year}
  {2018})}\BibitemShut {NoStop}%
\bibitem [{\citenamefont {Artrith}\ \emph {et~al.}(2017)\citenamefont
  {Artrith}, \citenamefont {Urban},\ and\ \citenamefont {Ceder}}]{artrith2017}%
  \BibitemOpen
  \bibfield  {author} {\bibinfo {author} {\bibfnamefont {N.}~\bibnamefont
  {Artrith}}, \bibinfo {author} {\bibfnamefont {A.}~\bibnamefont {Urban}},\
  and\ \bibinfo {author} {\bibfnamefont {G.}~\bibnamefont {Ceder}},\ }\bibfield
   {title} {\bibinfo {title} {Efficient and accurate machine-learning
  interpolation of atomic energies in compositions with many species},\ }\href
  {https://doi.org/10.1103/PhysRevB.96.014112} {\bibfield  {journal} {\bibinfo
  {journal} {Physical Review B}\ }\textbf {\bibinfo {volume} {96}},\ \bibinfo
  {pages} {014112} (\bibinfo {year} {2017})}\BibitemShut {NoStop}%
\bibitem [{\citenamefont {Gastegger}\ \emph {et~al.}(2018)\citenamefont
  {Gastegger}, \citenamefont {Schwiedrzik}, \citenamefont {Bittermann},
  \citenamefont {Berzsenyi},\ and\ \citenamefont {Marquetand}}]{gastegger2018}%
  \BibitemOpen
  \bibfield  {author} {\bibinfo {author} {\bibfnamefont {M.}~\bibnamefont
  {Gastegger}}, \bibinfo {author} {\bibfnamefont {L.}~\bibnamefont
  {Schwiedrzik}}, \bibinfo {author} {\bibfnamefont {M.}~\bibnamefont
  {Bittermann}}, \bibinfo {author} {\bibfnamefont {F.}~\bibnamefont
  {Berzsenyi}},\ and\ \bibinfo {author} {\bibfnamefont {P.}~\bibnamefont
  {Marquetand}},\ }\bibfield  {title} {\bibinfo {title}
  {{{wACSF}}---{{Weighted}} atom-centered symmetry functions as descriptors in
  machine learning potentials},\ }\href {https://doi.org/10.1063/1.5019667}
  {\bibfield  {journal} {\bibinfo  {journal} {The Journal of Chemical Physics}\
  }\textbf {\bibinfo {volume} {148}},\ \bibinfo {pages} {241709} (\bibinfo
  {year} {2018})}\BibitemShut {NoStop}%
\bibitem [{\citenamefont {Huo}\ and\ \citenamefont {Rupp}(2022)}]{huo2022a}%
  \BibitemOpen
  \bibfield  {author} {\bibinfo {author} {\bibfnamefont {H.}~\bibnamefont
  {Huo}}\ and\ \bibinfo {author} {\bibfnamefont {M.}~\bibnamefont {Rupp}},\
  }\bibfield  {title} {\bibinfo {title} {Unified representation of molecules
  and crystals for machine learning},\ }\href
  {https://doi.org/10.1088/2632-2153/aca005} {\bibfield  {journal} {\bibinfo
  {journal} {Machine Learning: Science and Technology}\ }\textbf {\bibinfo
  {volume} {3}},\ \bibinfo {pages} {045017} (\bibinfo {year}
  {2022})}\BibitemShut {NoStop}%
\bibitem [{\citenamefont {Darby}\ \emph {et~al.}(2022)\citenamefont {Darby},
  \citenamefont {Kermode},\ and\ \citenamefont {Cs{\'a}nyi}}]{darby2022}%
  \BibitemOpen
  \bibfield  {author} {\bibinfo {author} {\bibfnamefont {J.~P.}\ \bibnamefont
  {Darby}}, \bibinfo {author} {\bibfnamefont {J.~R.}\ \bibnamefont {Kermode}},\
  and\ \bibinfo {author} {\bibfnamefont {G.}~\bibnamefont {Cs{\'a}nyi}},\
  }\bibfield  {title} {\bibinfo {title} {Compressing local atomic neighbourhood
  descriptors},\ }\href {https://doi.org/10.1038/s41524-022-00847-y} {\bibfield
   {journal} {\bibinfo  {journal} {npj Computational Materials}\ }\textbf
  {\bibinfo {volume} {8}},\ \bibinfo {pages} {166} (\bibinfo {year}
  {2022})}\BibitemShut {NoStop}%
\bibitem [{\citenamefont {Mazitov}\ \emph {et~al.}(2024)\citenamefont
  {Mazitov}, \citenamefont {Springer}, \citenamefont {Lopanitsyna},
  \citenamefont {Fraux}, \citenamefont {De},\ and\ \citenamefont
  {Ceriotti}}]{mazitov2024}%
  \BibitemOpen
  \bibfield  {author} {\bibinfo {author} {\bibfnamefont {A.}~\bibnamefont
  {Mazitov}}, \bibinfo {author} {\bibfnamefont {M.~A.}\ \bibnamefont
  {Springer}}, \bibinfo {author} {\bibfnamefont {N.}~\bibnamefont
  {Lopanitsyna}}, \bibinfo {author} {\bibfnamefont {G.}~\bibnamefont {Fraux}},
  \bibinfo {author} {\bibfnamefont {S.}~\bibnamefont {De}},\ and\ \bibinfo
  {author} {\bibfnamefont {M.}~\bibnamefont {Ceriotti}},\ }\bibfield  {title}
  {\bibinfo {title} {Surface segregation in high-entropy alloys from alchemical
  machine learning},\ }\href {https://doi.org/10.1088/2515-7639/ad2983}
  {\bibfield  {journal} {\bibinfo  {journal} {Journal of Physics: Materials}\
  }\textbf {\bibinfo {volume} {7}},\ \bibinfo {pages} {025007} (\bibinfo {year}
  {2024})}\BibitemShut {NoStop}%
\bibitem [{\citenamefont {Lopanitsyna}\ \emph {et~al.}(2023)\citenamefont
  {Lopanitsyna}, \citenamefont {Fraux}, \citenamefont {Springer}, \citenamefont
  {De},\ and\ \citenamefont {Ceriotti}}]{lopanitsyna2023}%
  \BibitemOpen
  \bibfield  {author} {\bibinfo {author} {\bibfnamefont {N.}~\bibnamefont
  {Lopanitsyna}}, \bibinfo {author} {\bibfnamefont {G.}~\bibnamefont {Fraux}},
  \bibinfo {author} {\bibfnamefont {M.~A.}\ \bibnamefont {Springer}}, \bibinfo
  {author} {\bibfnamefont {S.}~\bibnamefont {De}},\ and\ \bibinfo {author}
  {\bibfnamefont {M.}~\bibnamefont {Ceriotti}},\ }\bibfield  {title} {\bibinfo
  {title} {Modeling high-entropy transition metal alloys with alchemical
  compression},\ }\href {https://doi.org/10.1103/PhysRevMaterials.7.045802}
  {\bibfield  {journal} {\bibinfo  {journal} {Physical Review Materials}\
  }\textbf {\bibinfo {volume} {7}},\ \bibinfo {pages} {045802} (\bibinfo {year}
  {2023})}\BibitemShut {NoStop}%
\bibitem [{\citenamefont {Thomas}\ \emph {et~al.}(2021)\citenamefont {Thomas},
  \citenamefont {Natarajan},\ and\ \citenamefont {{Van der Ven}}}]{thomas2021}%
  \BibitemOpen
  \bibfield  {author} {\bibinfo {author} {\bibfnamefont {J.~C.}\ \bibnamefont
  {Thomas}}, \bibinfo {author} {\bibfnamefont {A.~R.}\ \bibnamefont
  {Natarajan}},\ and\ \bibinfo {author} {\bibfnamefont {A.}~\bibnamefont {{Van
  der Ven}}},\ }\bibfield  {title} {\bibinfo {title} {Comparing crystal
  structures with symmetry and geometry},\ }\href
  {https://doi.org/10.1038/s41524-021-00627-0} {\bibfield  {journal} {\bibinfo
  {journal} {npj Computational Materials}\ }\textbf {\bibinfo {volume} {7}},\
  \bibinfo {pages} {164} (\bibinfo {year} {2021})}\BibitemShut {NoStop}%
\bibitem [{\citenamefont {Laks}\ \emph {et~al.}(1992)\citenamefont {Laks},
  \citenamefont {Ferreira}, \citenamefont {Froyen},\ and\ \citenamefont
  {Zunger}}]{laks1992}%
  \BibitemOpen
  \bibfield  {author} {\bibinfo {author} {\bibfnamefont {D.~B.}\ \bibnamefont
  {Laks}}, \bibinfo {author} {\bibfnamefont {L.~G.}\ \bibnamefont {Ferreira}},
  \bibinfo {author} {\bibfnamefont {S.}~\bibnamefont {Froyen}},\ and\ \bibinfo
  {author} {\bibfnamefont {A.}~\bibnamefont {Zunger}},\ }\bibfield  {title}
  {\bibinfo {title} {Efficient cluster expansion for substitutional systems},\
  }\href {https://doi.org/10.1103/PhysRevB.46.12587} {\bibfield  {journal}
  {\bibinfo  {journal} {Physical Review B}\ }\textbf {\bibinfo {volume} {46}},\
  \bibinfo {pages} {12587} (\bibinfo {year} {1992})}\BibitemShut {NoStop}%
\bibitem [{\citenamefont {Bl{\"o}chl}(1994)}]{blochl1994}%
  \BibitemOpen
  \bibfield  {author} {\bibinfo {author} {\bibfnamefont {P.~E.}\ \bibnamefont
  {Bl{\"o}chl}},\ }\bibfield  {title} {\bibinfo {title} {Projector
  augmented-wave method},\ }\href {https://doi.org/10.1103/PhysRevB.50.17953}
  {\bibfield  {journal} {\bibinfo  {journal} {Physical Review B}\ }\textbf
  {\bibinfo {volume} {50}},\ \bibinfo {pages} {17953} (\bibinfo {year}
  {1994})}\BibitemShut {NoStop}%
\bibitem [{\citenamefont {Kresse}\ and\ \citenamefont
  {Hafner}(1993)}]{kresse1993}%
  \BibitemOpen
  \bibfield  {author} {\bibinfo {author} {\bibfnamefont {G.}~\bibnamefont
  {Kresse}}\ and\ \bibinfo {author} {\bibfnamefont {J.}~\bibnamefont
  {Hafner}},\ }\bibfield  {title} {\bibinfo {title} {{\emph{Ab Initio}}
  molecular dynamics for liquid metals},\ }\href
  {https://doi.org/10.1103/PhysRevB.47.558} {\bibfield  {journal} {\bibinfo
  {journal} {Physical Review B}\ }\textbf {\bibinfo {volume} {47}},\ \bibinfo
  {pages} {558} (\bibinfo {year} {1993})}\BibitemShut {NoStop}%
\bibitem [{\citenamefont {Kresse}\ and\ \citenamefont
  {Furthm{\"u}ller}(1996{\natexlab{a}})}]{kresse1996}%
  \BibitemOpen
  \bibfield  {author} {\bibinfo {author} {\bibfnamefont {G.}~\bibnamefont
  {Kresse}}\ and\ \bibinfo {author} {\bibfnamefont {J.}~\bibnamefont
  {Furthm{\"u}ller}},\ }\bibfield  {title} {\bibinfo {title} {Efficient
  iterative schemes for {\emph{ab initio}} total-energy calculations using a
  plane-wave basis set},\ }\href {https://doi.org/10.1103/PhysRevB.54.11169}
  {\bibfield  {journal} {\bibinfo  {journal} {Physical Review B}\ }\textbf
  {\bibinfo {volume} {54}},\ \bibinfo {pages} {11169} (\bibinfo {year}
  {1996}{\natexlab{a}})}\BibitemShut {NoStop}%
\bibitem [{\citenamefont {Kresse}\ and\ \citenamefont
  {Furthm{\"u}ller}(1996{\natexlab{b}})}]{kresse1996a}%
  \BibitemOpen
  \bibfield  {author} {\bibinfo {author} {\bibfnamefont {G.}~\bibnamefont
  {Kresse}}\ and\ \bibinfo {author} {\bibfnamefont {J.}~\bibnamefont
  {Furthm{\"u}ller}},\ }\bibfield  {title} {\bibinfo {title} {Efficiency of
  ab-initio total energy calculations for metals and semiconductors using a
  plane-wave basis set},\ }\href {https://doi.org/10.1016/0927-0256(96)00008-0}
  {\bibfield  {journal} {\bibinfo  {journal} {Computational Materials Science}\
  }\textbf {\bibinfo {volume} {6}},\ \bibinfo {pages} {15} (\bibinfo {year}
  {1996}{\natexlab{b}})}\BibitemShut {NoStop}%
\bibitem [{\citenamefont {Ong}\ \emph {et~al.}(2013)\citenamefont {Ong},
  \citenamefont {Richards}, \citenamefont {Jain}, \citenamefont {Hautier},
  \citenamefont {Kocher}, \citenamefont {Cholia}, \citenamefont {Gunter},
  \citenamefont {Chevrier}, \citenamefont {Persson},\ and\ \citenamefont
  {Ceder}}]{ong2013}%
  \BibitemOpen
  \bibfield  {author} {\bibinfo {author} {\bibfnamefont {S.~P.}\ \bibnamefont
  {Ong}}, \bibinfo {author} {\bibfnamefont {W.~D.}\ \bibnamefont {Richards}},
  \bibinfo {author} {\bibfnamefont {A.}~\bibnamefont {Jain}}, \bibinfo {author}
  {\bibfnamefont {G.}~\bibnamefont {Hautier}}, \bibinfo {author} {\bibfnamefont
  {M.}~\bibnamefont {Kocher}}, \bibinfo {author} {\bibfnamefont
  {S.}~\bibnamefont {Cholia}}, \bibinfo {author} {\bibfnamefont
  {D.}~\bibnamefont {Gunter}}, \bibinfo {author} {\bibfnamefont {V.~L.}\
  \bibnamefont {Chevrier}}, \bibinfo {author} {\bibfnamefont {K.~A.}\
  \bibnamefont {Persson}},\ and\ \bibinfo {author} {\bibfnamefont
  {G.}~\bibnamefont {Ceder}},\ }\bibfield  {title} {\bibinfo {title} {Python
  {{Materials Genomics}} (pymatgen): {{A}} robust, open-source python library
  for materials analysis},\ }\href
  {https://doi.org/10.1016/j.commatsci.2012.10.028} {\bibfield  {journal}
  {\bibinfo  {journal} {Computational Materials Science}\ }\textbf {\bibinfo
  {volume} {68}},\ \bibinfo {pages} {314} (\bibinfo {year} {2013})}\BibitemShut
  {NoStop}%
\end{thebibliography}%

\end{document}